\documentclass[runningheads]{llncs}

\usepackage{amsmath}
\usepackage{booktabs} 
\usepackage{caption} 
\usepackage{subcaption} 
\usepackage{graphicx}
\usepackage{pgfplots}
\usepackage[all]{nowidow}
\usepackage[utf8]{inputenc}
\usepackage{tikz}
\usetikzlibrary{er,positioning,bayesnet}
\usepackage{multicol}
\usepackage{algpseudocode,algorithm,algorithmicx}
\usepackage{natbib}
\usepackage{hyperref}
\usepackage{cleveref}
\usepackage[inline]{enumitem} 
\definecolor{blue}{HTML}{1F77B4}
\definecolor{orange}{HTML}{FF7F0E}
\definecolor{green}{HTML}{2CA02C}

\pgfplotsset{compat=1.14}

\usepackage{a4wide}

\begin{document}

\title{Minimizing the polarization leakage of geometric-phase coronagraphs with multiple grating pattern combinations}

\author{David S. Doelman\inst{1,*}, Emiel H. Por\inst{1}, Garreth Ruane\inst{2}, Michael J. Escuti\inst{3}, Frans Snik\inst{1}}

\institute{Leiden Observatory, Leiden University, P.O. Box 9513, 2300 RA Leiden, The Netherlands
\email{*doelman@strw.leidenuniv.nl}\\ \and Jet Propulsion Laboratory, California Institute of Technology, 4800 Oak Grove Drive, Pasadena, CA, USA, 91109 \\ \and Department of Electrical and Computer Engineering, North Carolina State University, Raleigh, NC 27695, USA}

\titlerunning{Multiple-grating liquid-crystal coronagraphs}

\maketitle       

\begin{abstract}
The design of liquid-crystal diffractive phase plate coronagraphs for ground-based and space-based high-contrast imaging systems is limited by the trade-off between spectral bandwidth and polarization leakage. 
We demonstrate that by combining phase patterns with a polarization grating (PG) pattern directly followed by one or several separate PGs, we can suppress the polarization leakage terms by additional orders of magnitude by diffracting them out of the beam. 
\textcolor{black}{Using two PGs composed of a single-layer liquid crystal structure in the lab, we demonstrate a leakage suppression of more than an order of magnitude over a bandwidth of 133 nm centered around 532 nm. At this center wavelength we measure a leakage suppression of three orders of magnitude.}
Furthermore, simulations indicate that a combination of two multi-layered liquid-crystal PGs can suppress leakage to $<10^{-5}$ for 1-2.5 $\mu$m and $<10^{-10}$ for 650-800 nm.
We introduce multi-grating solutions with three or more gratings that can be designed to have no separation of the two circular polarization states, and offer even deeper suppression of polarization leakage. 
We present simulations of a triple-grating solution that has $<10^{-10}$ leakage on the first Airy ring from 450 nm to 800 nm. 
We apply the double-grating concept to the Vector-Vortex coronagraph of charge 4, and demonstrate in the lab that polarization leakage no longer limits the on-axis suppression for ground-based contrast levels. 
Lastly, we report on the successful installation and first-light results of a double-grating vector Apodizing Phase Plate pupil-plane coronagraph installed at the Large Binocular Telescope.
We discuss the implications of these new coronagraph architectures for high-contrast imaging systems on the ground and in space.
\end{abstract}

\section{Introduction}
\label{sec:int}

Direct imaging and spectral characterization of exoplanets constitutes an exciting yet immense challenge, set by the small angular separation of the exoplanet to the star and their extreme difference in brightness. 
\textcolor{black}{For a solar-system analogue at 10 parsec observed at visible wavelengths, an earth-twin has a separation of 100 mas and a contrast ratio of $\sim10^{-10}$.}
More massive self-luminous exoplanets and brown dwarfs in young systems have contrast ratios in the near-infrared that are less extreme, ${\sim}10^{-5}$ for HR 8799b,c,d,e \citep{marois2008direct}. 
Detecting and characterizing these young planets with direct imaging, already requires a large-aperture telescope, high-precision wavefront sensing and correction, a coronagraph with a small inner working angle to suppress the stellar diffraction halo by masking or diffracting stellar light, advanced observational and data-reduction strategies (e.g.~angular differential imaging (ADI) and spectral differential imaging (SDI)), and intricate optical measurement systems, such as integral field spectrographs (IFSs) and polarimetry \citep{marois2006angular,racine1999speckle,ruane2018review,jovanovic2018review,snik2018review}. 

\textcolor{black}{The unprecedented spatial resolution that the Extremely Large Telescopes (ELTs) currently under construction will offer will not only allow us to detect more exoplanets, but also characterize exoplanets found either by direct imaging or by other methods.}
Time-resolved, spectral and polarimetric measurements will provide crucial constraints on planet formation, atmospheric properties and eventually habitability and even habitation.
From the ground the focus will be on habitable planets orbiting M-dwarfs as their contrast is more favorable, but a large telescope aperture is necessary to resolve their angular separation. 
A next generation of space telescopes with extreme contrast performance is necessary to observe rocky planets in the habitable zones around sun-like stars.
Direct exoplanet characterization requires broadband observations to capture many photons and enable spectroscopy, and, therefore, coronagraph designs have to accommodate large spectral bandwidths. 
In addition to being able to spectrally characterize exoplanets, other advantages that broadband coronagraphs offer include increased observing flexibility and efficiency, diagnostics for wavelength-dependent instrument calibration, and spectral differential methods \citep{SparksFord2002,Snellen2015}. 
Even current ground-based high-contrast imaging systems could benefit from an enhanced broadband coronagraph implementation (provided sufficient wavefront control), as they all have a coronagraph in the common path and split up to different wavelength channels afterwards \citep{macintoshGPI,SCExAO,hinz2016overview,males2018magao,beuzit2019sphere}. 
Many coronagraphs installed in these systems are optimized for a single observing band, and perform suboptimal when used in a broad spectral band or at different wavelengths.

\subsection{Diffractive phase plate coronagraphs}

A promising type of coronagraph for broadband observations with small inner working angles is the diffractive phase plate coronagraph. 
Phase-only solutions often provide the most optimal combination of inner working angle and throughput, both for focal-plane coronagraphs and pupil-plane coronagraphs \citep{mawet2009optical,por2017optimal}.
Classical phase plates are chromatic by nature, but the so-called geometric phase \citep{Escuti:16} allows for the design and manufacturing of strictly achromatic devices, as the phase it applies only depends on the orientation pattern of the fast axis of a (optically flat) retarder.
A geometric-phase coronagraph has two important properties: the coronagraphic mask applies a phase pattern that is independent of wavelength and the fraction of light that acquires this phase depends on the retardance of the mask.
The first property demonstrates why geometric-phase coronagraphs are excellent candidates for broadband observations, while the second property can represent a major technical limitation. 
When the retardance of a mask deviates from half-wave, a fraction of light does not acquire the geometric phase, and thus produces a regular non-coronagraphic PSF. 
In this paper, we will refer to this fraction of light that does not acquire the geometric phase as ``polarization leakage''. Polarization leakage is one of the factors limiting the performance \citep{mawet2009vector,ruane2019scalar}.
Most geometric phase elements (including coronagraphs) are manufactured with patterned birefringent liquid crystals.
With a direct-write system \citep{Miskiewicz2014}, any geometric-phase pattern can be applied with $\sim$1 $\mu$m resolution through an alignment layer that is deposited on any flat substrate.
This orientation pattern imposed on the fast axis of consecutive liquid-crystal layers, which can be tuned to have half-wave retardance at a single wavelength (single-layer zero-order retardance), or to have close-to-half-wave retardance over a large spectral bandwidth by combining two or more liquid-crystal layers with different thicknesses and material properties like birefringence and twist \citep{Komanduri2013}.\\
Two types of geometric-phase coronagraphs have become popular: the Vector Vortex Coronagraph (VVC) and the vector Apodizing Phase Plate (vAPP).
\subsection{The Vector Vortex Coronagraph}
The VVC is one realization of the optical vortex coronagraph, a focal-plane coronagraph  \citep{foo2005optical} that diffracts all on-axis light out of the beam onto the Lyot stop for an unobstructed circular telescope aperture. 
The VVC has a phase mask in the focal plane and an amplitude mask (Lyot stop) in the downstream pupil plane to block the diffracted on-axis light. 
The phase mask has an azimuthal phase ramp of $2 \pi n$, where \textcolor{black}{$n$} is an integer indicating the ``charge'' of the coronagraph. 
\textcolor{black}{A higher charge offer larger resilience against optical aberrations including tip/tilt errors, but also increases the inner working angle \cite{ruane2017performance}}.
\textcolor{black}{Note that the VVC is very suitable for broadband implementation as its phase pattern is scale-invariant and therefore wavelength scaling of the PSF theoretically does not influence performance}.
The VVC has been installed on many telescopes and has been successful in detecting exoplanets and brown dwarfs \citep{serabyn2010image,absil2016three,serabyn2017wm}. 
The polarization leakage of a VVC mask does not limit the reached contrast for these ground-based observations in single spectral bands for which the masks were optimized.
Instead, the systems are limited by non-common path aberrations and adaptive optics residuals. 
However, for ground-based simultaneous multi-band observations \textcolor{black}{at $\sim$ 10$^{-5}$ raw contrast, limited by the Adaptive Optics residuals,} and space-based single-band observations \textcolor{black}{at $\sim$ 10$^{-10}$ raw contrast}, the leakage is still a limiting the attainable contrast.
\subsection{The vector Apodizing Phase Plate coronagraph}
\textcolor{black}{The vector-Apodizing Phase Plate (vAPP) is a version of the Apodizing Phase Plate (APP) coronagraph, a single-optic pupil-plane coronagraph \citep{Codona2006,Kenworthy2007}}.
An APP modifies the phase in the pupil plane to create regions in the PSF where the star light is suppressed, so-called dark zones. 
As opposite circular polarization states create a PSF with a dark hole on opposite sides, the vAPP implements circular polarization splitting that produces two complementary dark holes.
A truly broadband vAPP is obtained in combination with a Wollaston prism and a quarter-wave plate, but the non-coronagraphic PSFs corresponding to various polarization leakage terms degrade the contrast in the dark holes \citep{Snik2012,Otten2014a}.
The polarization leakage can be separated from the coronagraphic PSFs by adding a grating pattern (= phase tilt) to the phase pattern, i.e.~the ``grating-vAPP'' (gvAPP) \citep{Otten2014}. 
At present, one grating-vAPP is installed in the MagAO/Clio2 instrument at the Magellan Clay telescope (2--5 $\mu$m) \citep{Otten2017}, one at the MagAO-X instrument \citep{millervapp2019} (550-1100 nm) at the same telescope, one in the SCExAO instrument at the Subaru telescope (1--2.5 $\mu$m) \citep{Doelman2017}, and several other gvAPPs are being designed/manufactured/commissioned at the moment.
The grating introduces significant wavelength smearing for the two coronagraphic PSFs, reducing the simultaneous bandwidth of the gvAPP for standard imaging. 
This implementation is still applicable to integral-field spectroscopy, but the overall PSF structure is generally large in comparison with the typically limited field-of-view of an IFS.
In addition, half of the exoplanet light observed with a gvAPP with a D-shaped dark zone ends up on the bright-side of PSF, reducing the effective planet throughput.
\textcolor{black}{This can be negated by using phase solutions for 360$^\circ$ donut-shaped dark holes. However, these solutions are still limited by wavelength smearing when using a grating or on-axis leakage when used without a grating.}
Furthermore, the inner working angle is larger than for 2$\times$180$^\circ$ solutions.
\subsection{Leakage mitigation strategies}
Current leakage mitigation strategies have been applied mostly for the VVC, as the grating-vAPP is very resistant against leakage. For the VVC, previous efforts to reduce or mitigate the leakage for the VVC mask can be split into four approaches.
The first approach involves the engineering of the device such that its retardance is half-wave to within the tolerances set within the operational spectral band.
For liquid-crystal devices this can be achieved by adding more layers.
These additional can be self-aligning \citep{Komanduri2013}, or separate structures that are aligned and fixed manually \citep{mawet2011taking,roberts2019overcoming}.
The latter approach has recently been shown to reduce the leakage to less than $<0.02\%$ for 10\% bandwidth \citep{roberts2019overcoming} and around $0.1\%$ for 20\% bandwidth \citep{serabyn2019vector}. 
Geometric phase patterns can also be imposed through form birefringence through sub-wavelength grating patterns, which for the charge-2 VVC leads to the implementation of the Annular Groove Phase Mask (AGPM) \citep{Mawet2005}.
The retardance of such a device can be achromatized by tuning the depth and shape of the sub-wavelength grooves, and by combining different materials \citep{Vargas2016}.
This technology can deliver polarization leakage of a few 10$^{-3}$ over mid-infrared bands where this technology is readily applied \citep{jolivet2019and}.\\
Even after applying these advanced manufacturing methods the polarization leakage is filtered for space-equivalent extreme contrast experiments, by means of sandwiching the VVC between ``circular polarizers'' consisting of a linear polarizer and a quarter-wave plate \citep{mawet2010,Snik2014a}.
This therefore constitutes a second method, although this implementation does complicate the optical configuration.
Moreover, the polarizers and quarter-wave plates need to be of high quality to offer additional leakage suppression of $\sim$10$^{-8}$ to reach the contrast necessary to directly detect Earth-twins at 10$^{-10}$ contrast in the visible.\\ \textcolor{black}{A third approach that has recently been revived in \cite{ruane2019scalar}, where they use a scalar vortex instead of a vector vortex. }
A scalar vortex obviously has no polarization leakage and incident light acquires the same phase shift regardless of polarization state. 
But the major challenge is to achromatizing such a mask by using two different glass types, and optimizing the Lyot stop diameter and two deformable mirror shapes.\\
The fourth approach is most similar to the multi-grating concept presented in this paper, and adds a grating to the vector vortex phase pattern to yield a ``forked grating''. 
The forked grating separates the leakage spatially from the vortex beam. 
Only one orthogonal circular polarization state is selected by blocking the leakage and the opposite state. 
The diffraction of the grating is compensated in a different plane. 
This approach uses either a spatial light modulator with a computer generated hologram \citep{leach2003observation}, binary amplitude mask \citep{Ruane:14,kanburapa2012white} or two volume phase holograms \citep{Mariyenko:05}. 
These approaches reach high diffraction efficiency but filter $>50\%$ of the light and require complex setups that are difficult to integrate in current setups. \\
In this paper, we present the multi-grating concept to minimize the effects of polarization leakage for both the VVC and vAPP coronagraphs. 
The fundamental principle of the multi-grating concept is that the main polarization leakage terms are removed through diffraction, while the coronagraphic PSFs are recombined on-axis in a single optic. 
The diffracted leakage terms are diffracted outside the beam or the useful field-of-view.
It is a simple, yet powerful upgrade that is easily implemented in current and future high-contrast imaging systems, both on the ground and in space.\\
In Sect~\ref{Sec:Theory} we explain the multi-grating principle and demonstrate in simulation a leakage suppression of many orders of magnitude. 
We present our lab validation of a double-grating element in Sect.~\ref{Sec:dglabresults}.
In Sect.~\ref{Sec:corosystem} we introduce the coronagraph architectures of both a multi-grating focal-plane coronagraph (VVC) and a double-grating pupil-plane coronagraph (vAPP).
We demonstrate the performance of a double-grating VVC in Sect.~\ref{Sec:DG_VVC}, and of a double-grating vAPP at the Large Binocular Telescope in Sect.~\ref{Sec:DG_LBT}.
We discuss the implications on systems level of these novel coronagraph architectures in Sect.~\ref{sec:systemimplications}. 

\section{Double-grating diffraction theory}
\label{Sec:Theory}
Both the VVC and the vAPP coronagraph are commonly manufactured as geometric phase holograms (GPHs) \citep{Kim2015,Escuti:16}. In this section we will look into the properties of GPHs and how the geometry of phase patterns can be used to remove leakage. We simulate the performance for several double-grating combinations consisting of multi-layer liquid-crystal GPHs, and show drastically reduced polarization  leakage  over  large  spectral  bands.  Lastly,  we  characterize  the  influence  of  grating separation.
\subsection{The geometric phase hologram}
A geometric phase hologram is a half-wave retarder with a spatially varying fast-axis orientation. Circularly polarized light propagating through a GPH acquires geometric phase (or Pancharatnam-Berry phase \citep{Pancharatnam1955,pancharatnam1956generalized,berry1984quantal,Berry1988}), that depends on the fast-axis orientation. The space-variant Jones matrix of such a retarder in the circular polarization basis is given by
\begin{equation}
    \textbf{M} = c_V \begin{bmatrix} 0 & \text{e}^{i2\chi(x,y)} \\ \text{e}^{-i2\chi(x,y)} & 0 \text{ } \end{bmatrix} + c_L \begin{bmatrix} 1 \text{ } & 0 \\ 0 \text{ } & 1 \end{bmatrix}.
    \label{eq:GPH}
\end{equation}
Here $\chi(x,y)$ is the spatially varying fast-axis orientation, and both $c_V$ and $c_L$ are parameters that depend on the retardance $\Delta \phi$ \citep{mawet2009optical,ruane2019scalar} and are given by
\begin{equation}
c_V = \sin{\frac{\Delta \phi}{2}},\text{  } c_L = -i \cos{\frac{\Delta \phi}{2}}.
\end{equation}
The first term in Eq. \ref{eq:GPH} describes the fraction of the light that acquires a geometric phase of
\begin{equation}
\Phi(x,y) = \pm 2\chi(x,y), 
\end{equation}
where the sign of the phase depends on the handedness of the incoming circular polarization. 
The second term describes the polarization beam, and is unaffected by the fast-axis orientation pattern.
When the retardance is perfectly half-wave, i.e. $c_V = 1$ and $c_L = 0$, the two basis circular polarization states are converted to
\begin{equation}
\mathbf{RC_{out}} = \textbf{M} \mathbf{RC_{in}} = \textbf{M} \begin{bmatrix} 1 \\ 0 \end{bmatrix} =  \begin{bmatrix} 0 \\ \text{e}^{-i\Phi(x,y)} \end{bmatrix},
\end{equation}
and, 
\begin{equation}
\mathbf{LC_{out}} = \textbf{M} \mathbf{LC_{in}} = \textbf{M} \begin{bmatrix} 0 \\ 1 \end{bmatrix} =  \begin{bmatrix} \text{e}^{i\Phi(x,y)} \\ 0 \end{bmatrix}.
\end{equation}
From these equations we derive four properties of the geometric phase hologram:
\begin{enumerate}
  \item A GPH applies geometric phase that only depends on the local fast-axis orientation and is therefore independent of wavelength.
  \item The applied phase has an opposite sign for opposite handedness of the incoming circular polarization state.
  \item A GPH flips the circular polarization state. 
  \item When the retardance deviates from half-wave, the diffraction efficiency decreases and polarization leakage emerges.
\end{enumerate}
Unpolarized light has no preferred state of polarization and contains on average equal amounts of left and right circular polarization. 
When unpolarized light travels through a GPH, we can define three distinct waves emerging from the GPH. 
They are called the primary (+), conjugate (-) and leakage (0) wave \citep{Hasman2002,Kim2015,Escuti:16}.
\textcolor{black}{They correspond to the two basis vectors in the circular polarization basis, \textbf{LC} and \textbf{RC}, that acquire phase with opposite sign and the polarization leakage.}
We define the diffraction efficiencies for these waves as follows
\begin{equation}
\eta_{+} = |\left<\mathbf{E_{in}}|\mathbf{LC}\right>|^2 \sin^2(\Delta \phi /2), = |\left<\mathbf{E_{in}}|\mathbf{LC}\right>|^2 |c_V|^2,
\label{eq:LCleak}
\end{equation}
\begin{equation}
\eta_{-} = |\left<\mathbf{E_{in}}|\mathbf{RC}\right>|^2 \sin^2(\Delta \phi /2), = |\left<\mathbf{E_{in}}|\mathbf{RC}\right>|^2 |c_V|^2,
\label{eq:RCleak}
\end{equation}
\begin{equation}
\eta_{0} = \cos^2(\Delta \phi /2) = |c_L|^2.
\label{eq:leak}
\end{equation} 
\textcolor{black}{Here  $\left<\mathbf{A}|\mathbf{B}\right>$ is the dot product of \textbf{A} and \textbf{B}. While the individual diffraction efficiency of the primary and conjugate wave depends on the input polarization state, the overall diffraction efficiency of a GPH, as used in this paper, is given by $|c_V|^2 = \eta_{+} + \eta_{-}$. 
The diffraction efficiency is therefore independent of polarization state. 
Note that it is equivalent to define $|c_V|^2 = \eta_{\pm}$ when assuming fully circularly polarized light.}
The retardance can be tuned to be close to half-wave over large bandwidths by adding multiple layers with different thickness and twist \citep{Komanduri2013,Escuti2016}. 
Efficiencies are typically $\sim$99\% over spectral bandwidths as large as an octave, which implies a polarization leakage of $\sim$1\%. \\
\textcolor{black}{Classically}, the fast-axis orientation pattern and the retardance of a GPH can be treated as two separate properties, i.e.~the polarization leakage can be minimized independent of the fast-axis orientation pattern. 
The multi-grating concept is different from this classical idea to minimize polarization leakage because it uses the geometry of the phase pattern instead of tuning the retardance more precisely by adding more liquid-crystal layers. 
\subsection{The double-grating element}
\begin{figure}[t]
    \centering
        \includegraphics[width=\textwidth ]{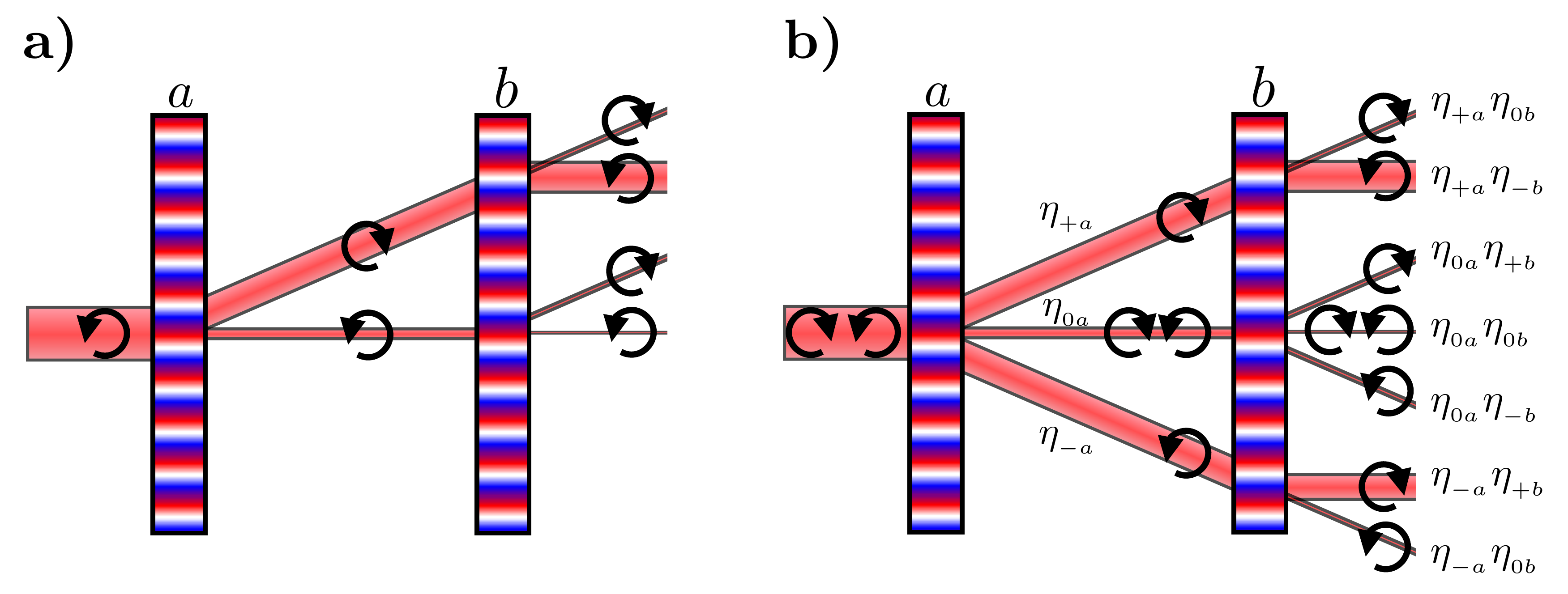}
        \caption{Schematic of the propagation of 100\% circularly polarized light (left) unpolarized light (right) through a double-grating element. The thickness of the beams correspond to the amount of light in each beam, assuming close-to $\lambda/2$-retardance for both polarization gratings. The main beams are diffracted back on-axis because of the polarization flip that occurs when propagating through a GPH.}
        \label{fig:DGleakagedir}
 \end{figure}

A polarization grating (PG) is a GPH with a uniformly rotating fast-axis orientation in a defined direction and a constant orientation in the orthogonal direction \citep{Oh2008,Packham2010}. 
The PG is therefore a GPH that applies a phase ramp in this defined direction. 
A PG splits two circular polarization states, diffracting them in opposite direction. 
One special property of a PG is that it has ``polarization memory'': contrary to regular gratings, a fully circularly polarized beam will in its entirety be diffracting in one particular order, barring polarization leakage. 
\textcolor{black}{This is a direct consequence of Eq. \ref{eq:LCleak} and \ref{eq:RCleak}.}\\
A double-grating element introduced in this paper, consists of two individual GPHs. 
The first GPH has an orientation pattern corresponding to a coronagraphic pattern, e.g a VVC or vAPP, with a coherently added polarization grating pattern. We call this GPH the coronagraphic polarization grating (CPG). 
The second GPH is a normal polarization grating that has the same pattern as the polarization grating that was added to the CPG. 
A schematic of how circularly polarized light propagates through a double-grating element is depicted in Fig. \ref{fig:DGleakagedir}. Here, the CPG is GPH $a$, and the PG is GPH $b$.
\\
When a beam with a single circular polarization state propagates through the CPG, the main beam is diffracted and a minor leakage term propagates on-axis. 
The PG diffracts the main beam back on-axis, as the PG patterns are equivalent, but the circular polarization state has been flipped by the CPG. 
The real benefit of the double-grating concept is demonstrated by what happens with the leakage term of the CPG. 
This leakage term is mostly diffracted out of the beam, such that the remaining on-axis leakage is further suppressed by a factor $\eta_{0 a}\eta_{0 b}$. 
For unpolarized light, both the primary and the conjugate beam of the CPG are diffracted back on-axis by the PG, given the polarization memory of PGs. 
In conclusion, a double-grating element suppresses leakage by diffracting most of the leakage off-axis while diffracting the main beam twice in opposite direction with high efficiency, keeping the main beam on-axis. \\
If the retardance of both the CPG and the PG are close to half-wave, the diffraction efficiencies $(\eta_{+ a,b} + \eta_{- a,b}) \approx 1$ and $\eta_{0 a,b} \ll 1$. 
As the leakage is always the product of the two leakage factors $\eta_{0 a}\eta_{0 b}$, the leakage for two identical liquid-crystal films is suppressed by orders of magnitude.
As an example, two polarization gratings with each $1\%$ leakage will have a on-axis polarization leakage as low as $10^{-4}$. 
We emphasize that this is achieved with phase pattern geometry and not by optimizing the manufacturing process to optimize the retardance of individual optics beyond the $\sim$1\% leakage performance offered by standard liquid-crystal techniques. \\
Combining gratings this way is only possible with polarization gratings, and no other type of grating, because PGs operate on circular polarization states and diffract into a single order. 
This polarization memory is key to make this setup highly efficient. \\
Without polarization splitting, the double-grating concept constrains the design of the phase pattern. This constraint comes from the property that a double-grating element introduces phase with opposite sign for both circular polarization states. 
Phases with opposite sign in the pupil plane correspond to PSFs in the focal plane that are point-symmetric mirror images. 
If the PSFs are not point-symmetric, both circular polarization states will contaminate each other.
This limits the use of the double-grating concept to point-symmetric patterns like the vortex phase, or $0-\pi$ phase patterns, e.g.~in the pupil plane to produce PSFs with 360$^\circ$ dark holes. 

\subsection{Simulated broadband performance of double-grating elements}
\begin{figure}[t]
    \centering        \includegraphics[width=0.8\textwidth, trim = 100 0 80 0 ]{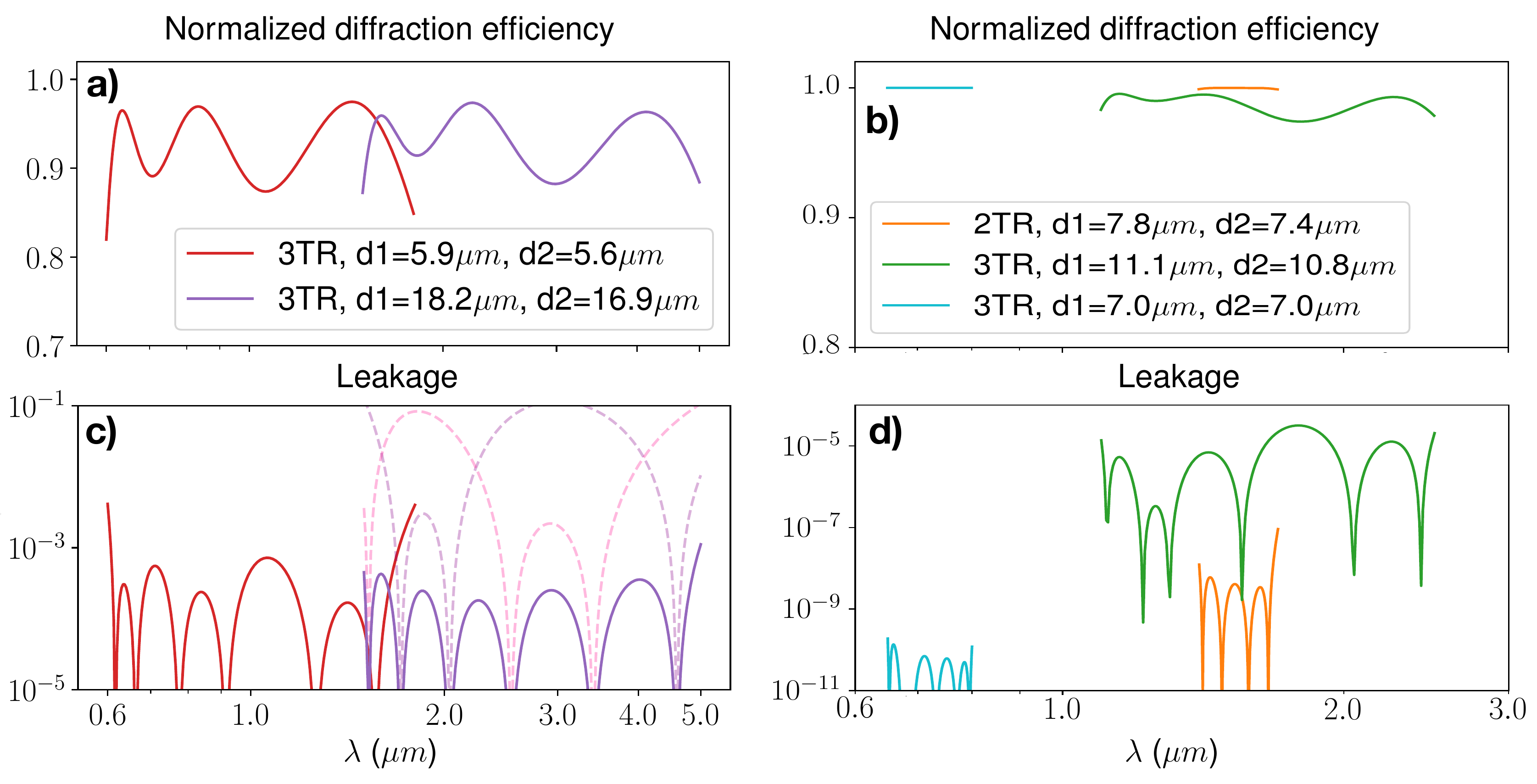}
        \caption{\textbf{Top:} Simulated diffraction efficiencies for five double-grating elements, with the average diffraction efficiency being higher than 90\%. The combined thickness of the layers are shown with d1 and d2, for the CPG and PG respectively. \textbf{Bottom:} Leakage as function of wavelength for the same double-grating elements. The dotted lines indicate the leakage of both PGs making up the double-grating element. These single PGs have different liquid-crystal recipes to minimize the total leakage.}
        \label{fig:DGleakagebb}
 \end{figure}

\textcolor{black}{
 We simulate the on-axis terms for different liquid-crystal recipes to demonstrate the broadband performance of a double-grating element. 
 Analogous to the first approach of minimizing polarization leakage in Sect.~\ref{sec:int}, we optimize the liquid-crystal recipes to minimize leakage for both the CPG and the PG in the double-grating element, \textcolor{black}{assuming a flat spectrum and fully circularly polarized light}.
 These recipes consist of multiple individual liquid-crystal layers, each with a thickness and chiral twist, that are arranged into a monolithic film, i.e. a multi-twist retarder (MTR)~ \citep{Komanduri2013,Kim2015,Escuti2016}. 
 We tune the retardance by adjusting the thickness and chiral twist for each layer in a two-layered retarder (2TR) or three-layered retarder (3TR). 
 The recipes of the CPG and the PG are optimized simultaneously using simulated annealing. In these simulations, the coronagraphic pattern is set to zero, such that only two PGs remain.}
 \textcolor{black}{We investigate solutions for three different regimes, each with different applications. 
 First, we optimize for extremely large bandwidths ($>90\%$) with moderate leakage suppression of $<10^{-3}$ that are capable of feeding multiple ground-based instrument arms. We show two solutions in the left panels of Fig.~\ref{fig:DGleakagebb}.
 Here we define the bandwidth as $\Delta \lambda/\lambda_0$, where $\Delta \lambda$ is the total bandwidth in nm and $\lambda_0$ is the central wavelength. 
 Second, we develop a recipe for good leakage suppression ($<10^{-5}$) for a bandwidth up to (90\%).
 Third, we explore possible recipes for small bandwidths (20\%) with extremely low leakage ($<10^{-8}$) that could be used for a single band in a space-based instrument. Solutions of the second and third regimes are shown in the right panels of Fig.~\ref{fig:DGleakagebb}}.
 
\textcolor{black}{
For the largest bandwidths, we optimize two double-grating recipes that are shown in \textcolor{black}{Fig.~\ref{fig:DGleakagebb}(a) and Fig.~\ref{fig:DGleakagebb}(c)}. 
These recipes have bandwidths of $100\%$ or more, the first one ranging from 0.6 $\mu$m to 1.8 $\mu$m (red) and the second one ranging from 1.5 $\mu$m to 5 $\mu$m (purple) with a leakage $<10^{-3}$.
Optimizing the recipes of the GPH and PG simultaneously yields non-trivial solutions where the two elements have different recipes. 
This is illustrated by the dashed lines that show the individual leakage terms for the PG and GPH. 
While the combined leakage is $<10^{-3}$ at all wavelengths, the individual recipe for the GPH or the PG can have as much as $10\%$ leakage for certain wavelengths, which is then compensated by the other element that has less $1\%$ leakage at that wavelength. 
This compensation is critical for extending the bandwidth beyond the broadband range that single-element 3TRs reach, which is typically up to $90\%$ for a leakage lower than $3\%$ \citep{Doelman2017}. 
A downside is that the increased leakage for single elements of the double-grating reduces diffraction efficiency. Therefore, the average combined diffraction efficiency is reduced to $93\%$ for these two designs. \\
For the second regime with wavelength ranges covered by a single ground-based instrument arm like an IFS, the goal is to suppress the leakage PSF below the speckle limit for extreme adaptive optics systems, which is around $10^{-5}$ in good circumstances. 
We optimize the double-grating recipes for the SCExAO \citep{SCExAO} instrument, operating between 1.1 $\mu$m and 2.5 $\mu$m combined with the integral-field spectrograph CHARIS \citep{CHARIS}.
The grating-vAPP that is currently installed in this instrument has an average leakage of $2\%$ over this bandwidth \citep{Doelman2017}.
For a vector vortex coronagraph this would limit the performance.
However, using the double-grating concept reduces the leakage by three additional orders of magnitude, going from $2\%$ to $<0.001\%$, while having a similar diffraction efficiency ($>97\%$), as shown with the 3TR recipe (green) in \textcolor{black}{Fig.~\ref{fig:DGleakagebb}(b) and Fig.~\ref{fig:DGleakagebb}(d).} \\
For space-based instruments, extremely low leakage is required. One advantage is that the bandwidth is limited, as the instantaneous wavefront correction with two deformable mirrors at these extreme contrast levels is limited to $\sim$20\% \citep{HabEx2019}.
As a first example, using two 2TRs for less complexity, it is possible to suppress the leakage from 1.5 $\mu$m to 1.8 $\mu$m (H-band, orange) to $10^{-8}$. In addition, we optimize for a wavelength range of 650-800 nm, as visible bands such as this one are planned for future space missions \citep{HabEx2019}. The transmission and leakage are shown in \textcolor{black}{Fig.~\ref{fig:DGleakagebb}(b) and Fig.~\ref{fig:DGleakagebb}(d)}, indicating that two three-layer stacks have an average polarization leakage of less than $10^{-10}$ and an average diffraction efficiency higher than $99.9\%$. 
This shows that a double-grating element can suppress the leakage to the required levels. 
Nevertheless, we introduce triple-grating solutions in Sect.\ref{Sec:multi-grating} to cover the entire visible spectral range with one coronagraph and enable extremely high contrast performance at any smaller band within that range.
Note that it may be sufficient to suppress the first Airy ring of the leakage PSF to $10^{-10}$, not the core itself. 
As the first Airy ring is typically at $\sim$ 10$^{-2}$ of the core intensity, this could reduce the complexity of liquid-crystal recipes that are used.
}

\subsection{The effect of non-zero grating separation}\label{sec:gratingseparation}

\begin{figure*}[t]
	\centering
	\includegraphics[width= \textwidth]{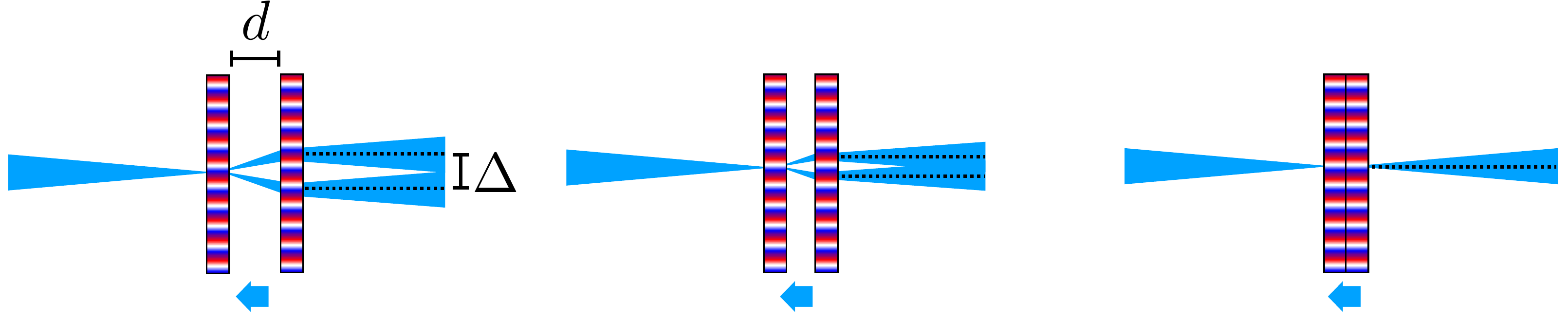}
		\caption{\textcolor{black}{Any finite distance, d, between two polarization gratings generates a shift, $\Delta$, between the primary and the conjugate beam. This separation decreases linearly with distance between the two gratings.}}
	     	\label{fig:splitfromdg}
\end{figure*}

From Fig.~\ref{fig:DGleakagedir} it is apparent that the polarization gratings split the beam in two separated main coronagraphic beams with opposite circular polarization. 
For the pupil-plane double-grating vAPP, a small shift of the two pupil on the order of a few micron is generally inconsequential, whereas for the focal-plane double-grating VVC this broadens or even splits up the off-axis PSF.
This effect is further elaborated in Fig.~\ref{fig:splitfromdg}.\\
We derive the beam shift by using the grating equation with $m=1$ to calculate the total lateral beam shift, $\Delta$, to be
\begin{equation}
\Delta = 2 d \tan \left(\arcsin\left(  \frac{\lambda}{P} \right)  \right).
\label{Eq:sep_v_sep}
\end{equation}
Here, $d$ is the separation between the plates, $P$ the period of the polarization grating, and the factor two comes from the two circular polarization states being diffracted in opposite direction.
For a pupil plane the effect is insignificant as $\Delta$ is orders of magnitude smaller than the pupil diameter.
However, the grating separation in the focal plane quickly becomes an issue, i.e. when $\Delta$ is a significant fraction of $\lambda/D$. 
We can minimize $\Delta$ by maximizing polarization grating period and minimizing the distance between the gratings. 
The optical system determines the maximum grating period, as the leakage needs to be diffracted out of the beam by a full pupil distance in the next pupil plane. 
Because the geometric phase is applied very locally \citep{Escuti:16}, the gratings can be very close together and still operate independently.
Moreover, the minimal distance of the gratings is set by the thickness of an optical adhesive that is used to glue both gratings together. 
Therefore, a finite shift is unavoidable. 
Even so, most high-contrast imaging systems have large F-numbers and in practice this shift is much less than the width of the point-spread function. 
As an example, we take an optical system with $\lambda = 1500nm$, a grating period of $30\mu m$, a separation given by a glue layer of 50 $\mu m$ and an F-number of 50, which gives a separation of 0.067 $\lambda/D$. 
For smaller F-numbers, significant wavelength-dependent splitting will occur for all the objects in the field, decreasing point-source detectability and astrometric and photometric accuracy.
\textcolor{black}{In Sect.~\ref{Sec:multi-grating} we introduce multi-grating solutions that solve this PSF splitting issue by design.}

\subsection{Multi-grating coronagraph architectures}\label{Sec:corosystem}
\begin{figure*}[h]
	\centering
	\includegraphics[width=\textwidth]{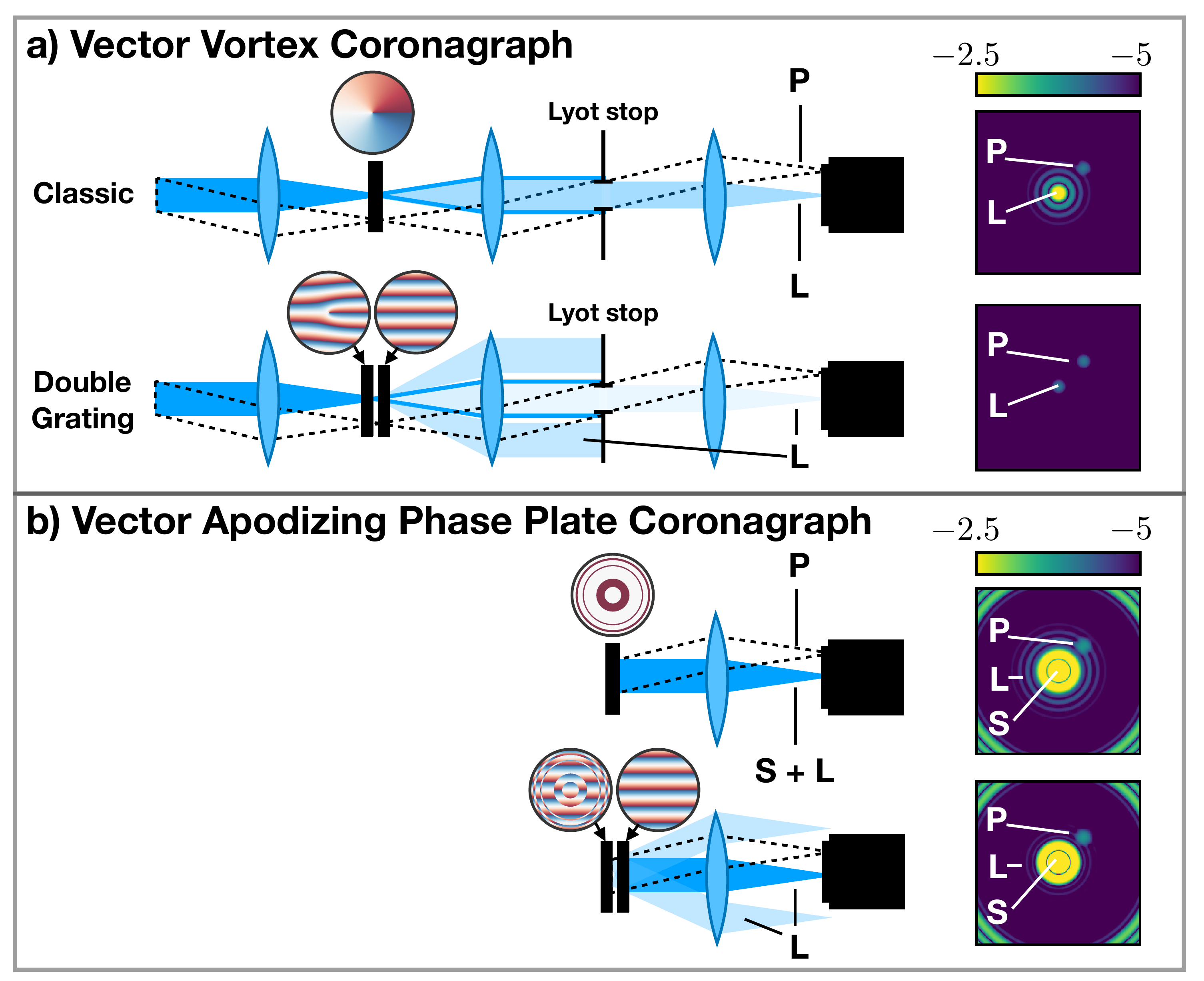}
		\caption{Comparison between the Vector Vortex coronagraph, a), and the vector Apodizing Phase Plate coronagraph, b), and their double-grating counterparts. The schematics show the optical paths of the diffracted leakage terms on a system level. The simulated point-spread functions demonstrate the effect of removing additional leakage when the double-grating coronagraphs are used. We annotate the planet PSF, P, the coronagraphic leakage, L, and the stellar PSF, S.}
	     	\label{fig:dgimplementation}
\end{figure*}

In this section, we discuss where the leakage terms are blocked for a double-grating implementation of both the VVC and the vAPP. 
Schematic drawings of coronagraphic systems with a focal-plane phase coronagraph like the VVC, and with a \textcolor{black}{pupil-plane} phase coronagraph like the vAPP are presented in Fig.~\ref{fig:dgimplementation} for a classical and double-grating configuration. 
The simulated point-spread functions include no aberration, a planet at a level of $10^{-4}$, and we assume a leakage of 1\% for each element.
 
For the VVC in Fig.~\ref{fig:dgimplementation}(a), most of the leakage in the classic configuration passes through the Lyot stop and is imaged onto the detector.
The double-grating VVC however, diffracts $\sim$99\% of the leakage term outside of the beam such that it is blocked by the Lyot stop, reducing the on-axis leakage to the level of $10^{-4}$.
\textcolor{black}{To diffract the non-coronagraphic leakage terms out of the beam, we require a grating period of $P < f\lambda/D$, where $\lambda$ is the wavelength, and $f$ and $D$ are the focal length and the diameter of the lens before the double-grating element, respectively. 
The leakage terms are non-coronagraphic pupils, and have minimal intensity outside of their diameter.}
\textcolor{black}{For leakage terms that have the coronagraphic pattern imprinted and therefore have light diffracted outside of the pupil plane, the separation needs to be larger, based on how much light is tolerable in the actual pupil of the main beam traveling through the Lyot stop.}\\
The regular (no-grating) vAPP with an annular dark hole has a leakage PSF that is imaged onto the same location as the apodized PSF.
This leakage term introduces extra light in the dark hole at the $10^{-4}$ level close to the inner edge of the dark zone at 3$\lambda/D$, as shown by the Airy rings present in the PSF of \textcolor{black}{Fig.~\ref{fig:dgimplementation}(b)}. 
The double-grating vAPP diffracts the leakage terms beyond the outer edge of the dark zone when the grating period $P<D/(2\Omega)$, where $D$ is the diameter of the pupil of the vAPP and $\Omega$ is the outer working angle in $\lambda/D$. 
Here we take into account that one of the two diffracted leakage terms is also apodized, see Fig.~\ref{fig:DGleakagedir}. 
\textcolor{black}{The double-grating version of both coronagraphs clearly reduce the signal of the leakage at the location of close-in companions. }

\section{Experimental results of double-grating implementations}
\label{Sec:dglabresults}
In the previous sections we derived two properties of the double-grating concept.
First, we mathematically showed that the on-axis polarization leakage is suppressed by additional orders of magnitude, and second, we derived that small grating separations do not generate significant polarization splitting for large F-numbers.
Here, we demonstrate these two properties with lab measurements. 
We characterize the double-grating performance with two polarization gratings, by measuring all leakage terms individually. 
Furthermore, we investigate the impact of grating separation on the polarization splitting by measuring the splitting as function of separation.

\subsection{Lab setup and double-grating manufacturing}
 
 \begin{figure*}[t]
	\centering
	\includegraphics[width= 0.8\textwidth, trim = 100 0 80 0]{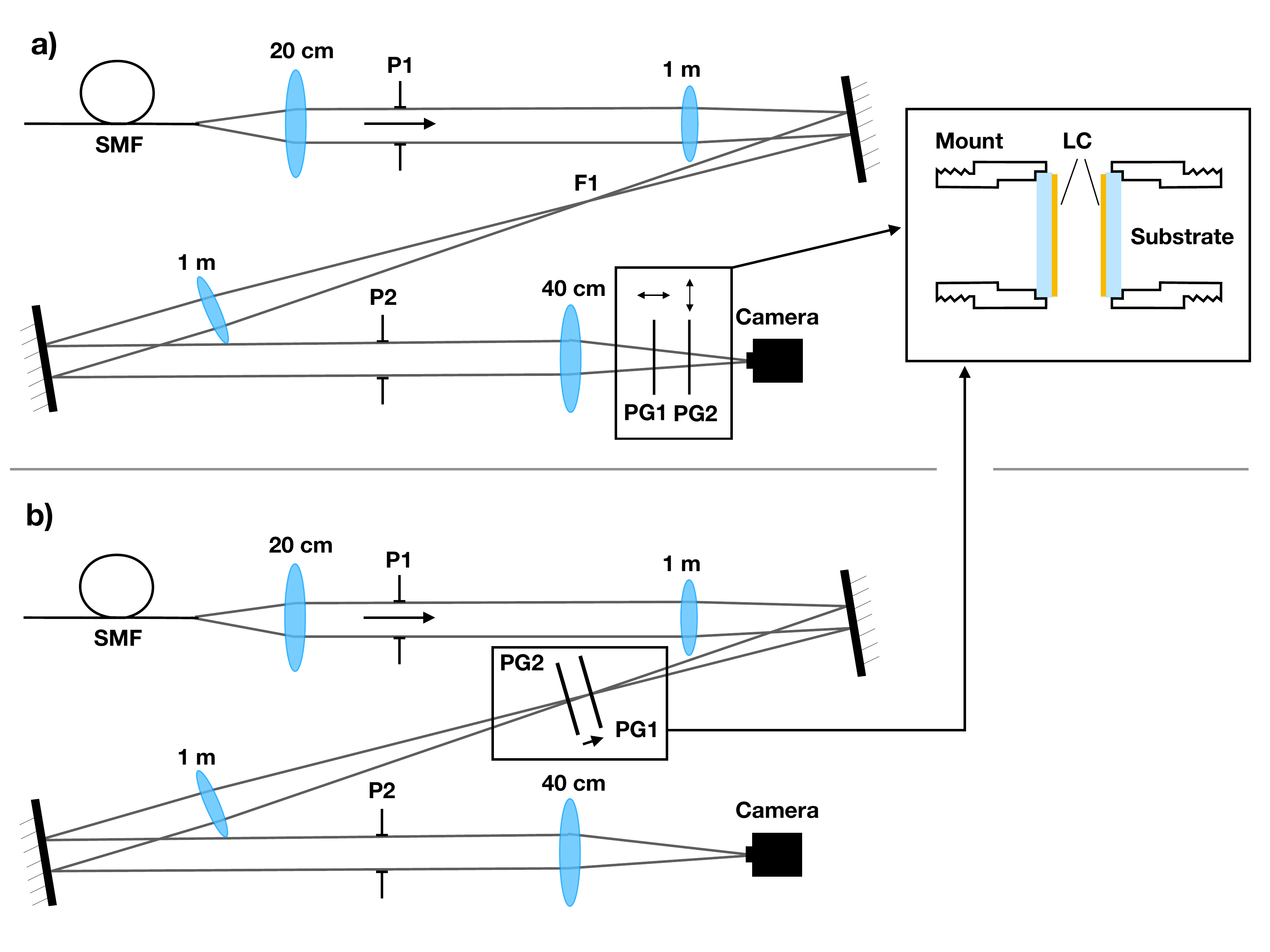}
		\caption{The two configurations of the lab setup used to measure the leakage as function of wavelength and PSF separation as function of grating separation respectively.
		}
	     	\label{fig:sep_setup}
\end{figure*}

We characterize the performance of the double-grating elements by placing two PGs in an imaging setup that is fed with a single-mode fiber. 
There are two configurations with different locations of the double-grating element.
These two configurations of the setup are shown in Fig. \ref{fig:sep_setup}.
We reimage the fiber onto the camera with an in-between focal plane.
The camera is a FLIR Chameleon3 2.8 MP mono camera operating at 12 bits.
We define a pupil plane before the double-grating element with an iris.
A second pupil plane contains the Lyot stop if the double-grating VVC is used. 
For both configurations, the single mode fiber has an injection unit for broadband light from an Ocean Optics HL 2000 halogen lamp and two Thorlabs laser diodes (CPS532a and CPS635S), operating at 532 nm and 635 nm. 
\textcolor{black}{For the halogen lamp we use Thorlabs FKB-VIS-10 narrowband filters (FWHM = 10 nm) spaced at 50 nm intervals between 500 nm and 800 nm. }
We use a polarizer before the first pupil plane (P1) to equalize the intensity in the two circular polarization states.
The intensity is equal because linear polarization can be decomposed in equal amounts of left and right circular polarization.
In the first configuration we measure the polarization leakage as function of wavelength.
The second configuration allows us to determine the PSF splitting for non-zero distance between the PGs and to evaluate the performance of the double-grating vector vortex coronagraph.\\
 We manufactured two identical PGs on two different substrates. 
 The PGs are printed on 2 inch glass (D263) windows with a thickness of 0.7 mm, have a period of 17 micron, a pixel size of 1 micron and are 10 mm by 10 mm in diameter.
 The PG pattern is written in a photo-alignment layer using the direct-write laser scanning system \citep{Miskiewicz2014,Kim2015} using a solid-state 355 nm laser. 
 The retardance of the liquid-crystal pattern is optimized using three sublayers of birefringent non-twisting self-aligning liquid-crystals to form a single thin film (1TR) that is cured with a 365 nm LED \citep{Komanduri2013}. 
 We measured the layer thickness to be approximately $1.1 \mu$m using an ellipsometer, corresponding to a half-wave retardance for 532 nm. 
\textcolor{black}{A more detailed description of the manufacturing of these PGs can be found in \cite{Doelman2019}}.
The two substrates are glued into a mount with a small groove, facing outward. 
The substrate extends beyond the edge of the mount.
This way, two liquid-crystal surfaces can touch when gently pressed together. 
The separation between the mounts can be regulated using translation stages. 

\subsection{Characterization of the double-grating polarization leakage}
\label{sec:dgleakage}
\begin{figure}[t]
    \centering
        \includegraphics[width= 0.8\textwidth, trim = 0 0 0 0 ]{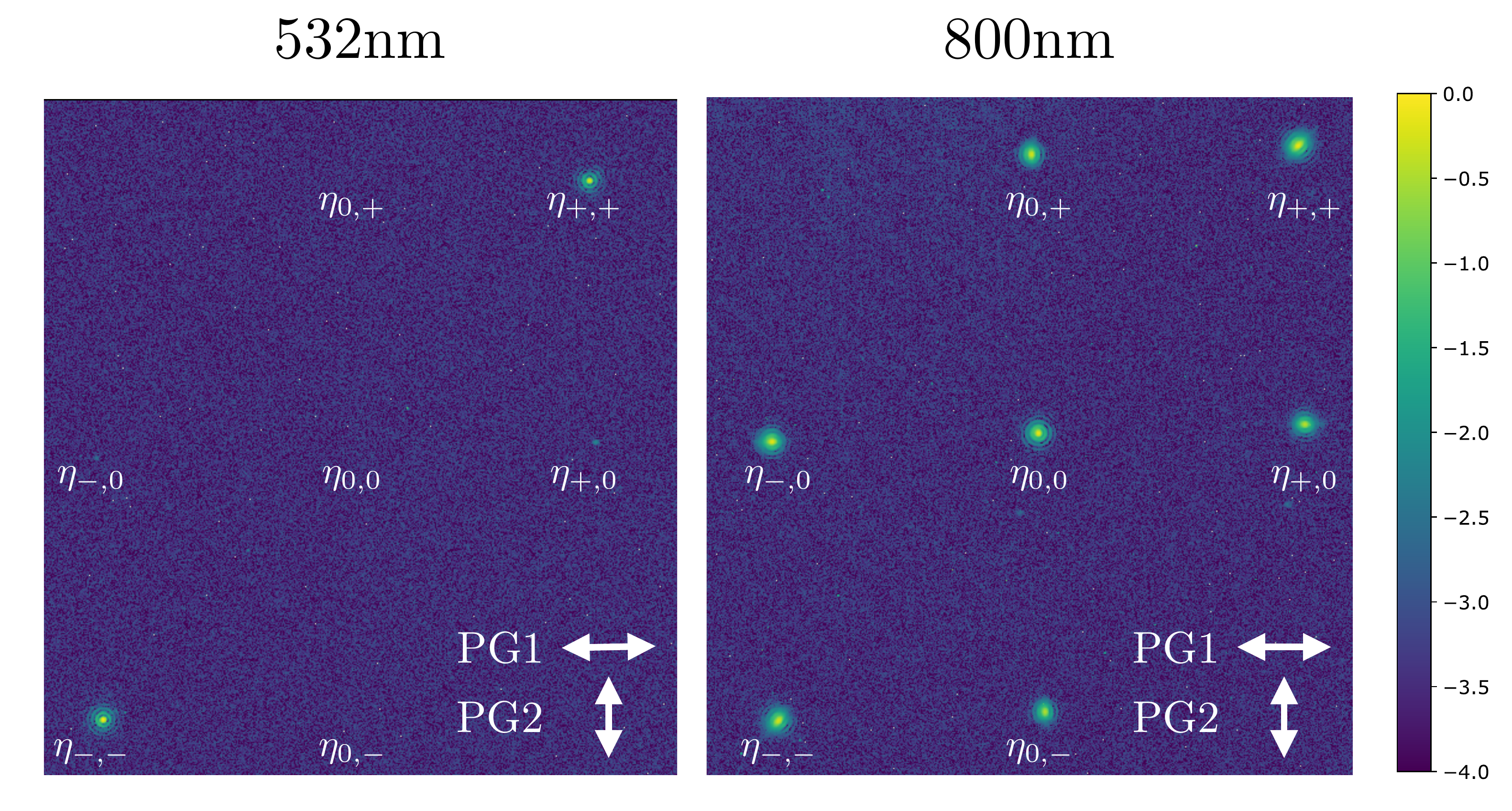}
        \caption{\textcolor{black}{Measured lab PSFs of a single source imaged through crossed polarization gratings that were optimized for 532 nm. The PGs are crossed to separate all leakage terms onto individual locations in the focal plane.} The spot identifiers, e.g. $\eta_{+,+}$, correspond to the primary beam with efficiency  $\eta_{+a}\eta_{+b}$. \textbf{Left:} The diffraction efficiency of each grating is optimized for 532 nm such that $99\%$ of the light is imaged on the combined first order of these two gratings. \textbf{Right:} The diffraction efficiency at 800 nm is less than $50\%$, creating all zero and first order diffraction spots of roughly equal brightness. Note that the absence of $\eta_{-,+}$ and $\eta_{+,-}$ is due to the polarization splitting of the first grating.}
        \label{fig:DGleakagefig}
 \end{figure}

To measure the double-grating polarization leakage, we use configuration a, with crossed polarization gratings. When the PGs are crossed, all leakage terms are imaged onto different locations in the focal plane. 
We separate the PGs by a small amount ($\sim 10$ mm), and increase the distance to the camera until the leakage PSFs are well separated in the focal plane.
The resulting images at 532 nm and 800 nm are shown in Fig.~\ref{fig:DGleakagefig}. 

\begin{figure}[t]
    \centering
        \includegraphics[width= 0.85\textwidth]{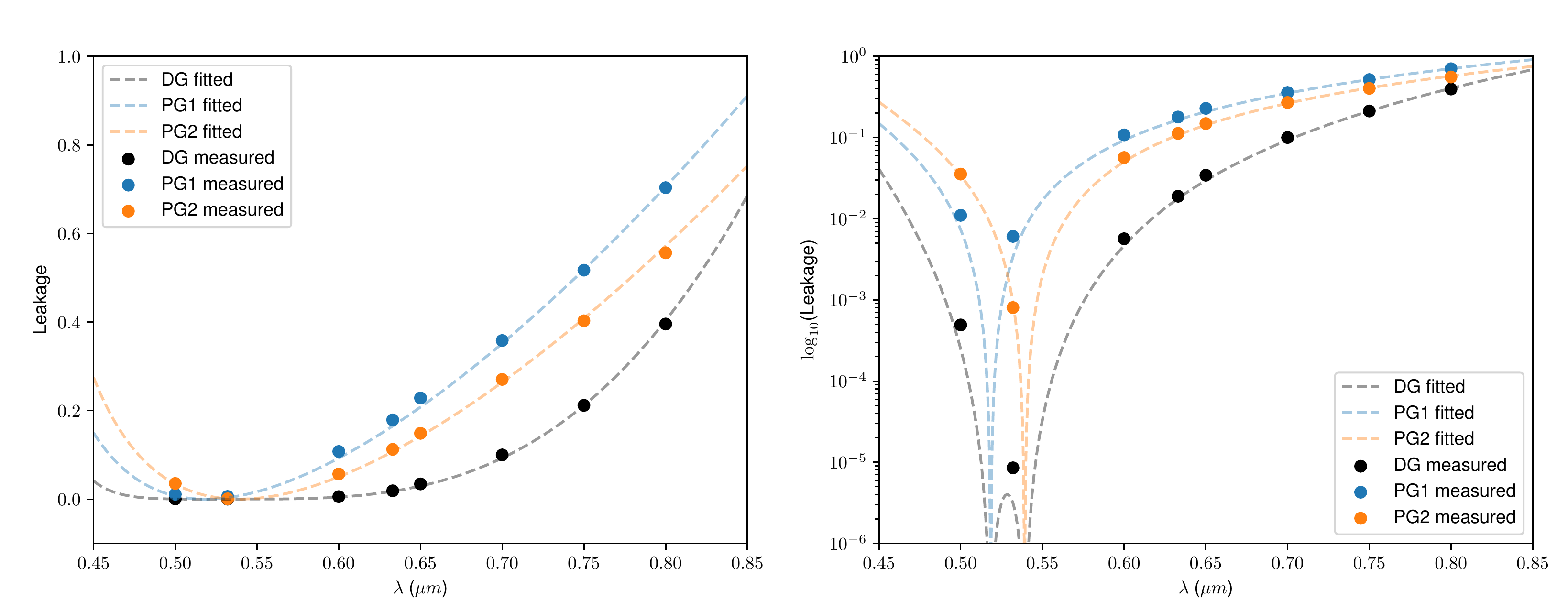}
        \caption{\textbf{Left:} \textcolor{black}{Lab data of the relative leakage intensity with fitted models as function of wavelength in linear scale.} \textbf{Right:} The same data and models in logarithmic scale. The relative double-grating leakage is less than $1\%$ over a bandwidth of 133 nm and less than $10^{-5}$ at 532 nm.}
        \label{fig:DGleakagewl}
    \centering
\end{figure} 
These images are used to reconstruct the individual diffraction efficiencies for both PGs using aperture photometry. 
\textcolor{black}{To remove transmission effects, we normalize the intensities on the total flux measured in all 8 spots for each wavelength.
The PGs are measured simultaneously in a stack, which means that we measure combinations, such as $\eta_{-,-} = \eta_{-,1} \eta_{-,2}$. Here $1$ and $2$ are used to distinguish between PG1 and PG2 respectively. }
To recover the diffraction efficiency of the first grating, $\eta_{-,1}$, from the image, we add three terms
\begin{equation}
\eta_{-,-} + \eta_{+,+} + \eta_{-,0} = \eta_{-,1} \eta_{-,2} + \eta_{+,1} \eta_{+,2} + \eta_{-,1} \eta_{0,2}.
\end{equation}
We use that $\eta_{+,a} = \eta_{-,a}$ for linear polarization and normalize the flux of the PSFs such that
\begin{equation}
\eta_{-,-} + \eta_{+,+} + \eta_{-,0} = \eta_{-,1} \left( \eta_{-,2} + \eta_{+,2} + \eta_{0,2} \right) = \eta_{-,1}.
\end{equation}
For high-contrast imaging we are interested in the power of the leakage terms of the gratings compared to the first-order diffraction efficiency. 
This relative intensity sets the contrast of the leakage term compared to a stellar PSF. 
We calculate these relative leakage intensities for both polarization gratings and the double-grating and these are plotted in linear scale (left) and logarithmic scale (right) in Fig.~\ref{fig:DGleakagewl}. 
We fit these relative leakage intensities for both polarization gratings using a Mueller matrix model of a twisted nematic cell with zero twist and varying thickness and fit the dispersion of the birefringence using a Cauchy expansion. 
The model for the double-grating leakage is the multiplication of the models for both PGs. 
The fitted models are in good agreement with both the measured individual leakage intensities and the combined double-grating leakage. 
The \textcolor{black}{fitted} layer thickness are 1.07 $\mu$m for PG1 and 1.14 $\mu$m for PG2. 
\textcolor{black}{The double-grating element reduces the leakage significantly over the full wavelength range and even by more than an order of magnitude between 500 nm and 633 nm.} Similarly, we infer from the models the bandwidth where the leakage is lower than $1\%$. The individual PGs reach a leakage suppression lower than $1\%$ over 46 nm bandwidth, compared to 150 nm for the double-grating. 
In addition, at 532 nm the double-grating leakage is lower than $1\times10^{-5}$, compared to $6\times10^{-3}$ or $7\times10^{-4}$ for PG1 and PG2 respectively.
These results show that by using two polarization gratings in succession it is possible to significantly reduce the polarization leakage in the on-axis beam.
\subsection{Separation of the double-grating PSF}

We measure the splitting from non-zero distance between the two PGs using configuration b shown in Fig.~\ref{fig:sep_setup}. 
We inject the 532 nm laser in a single mode fiber that is reimaged on the two polarization gratings with an F-number of 100. 
 \begin{figure*}[t]
	\centering
	\includegraphics[width= \textwidth]{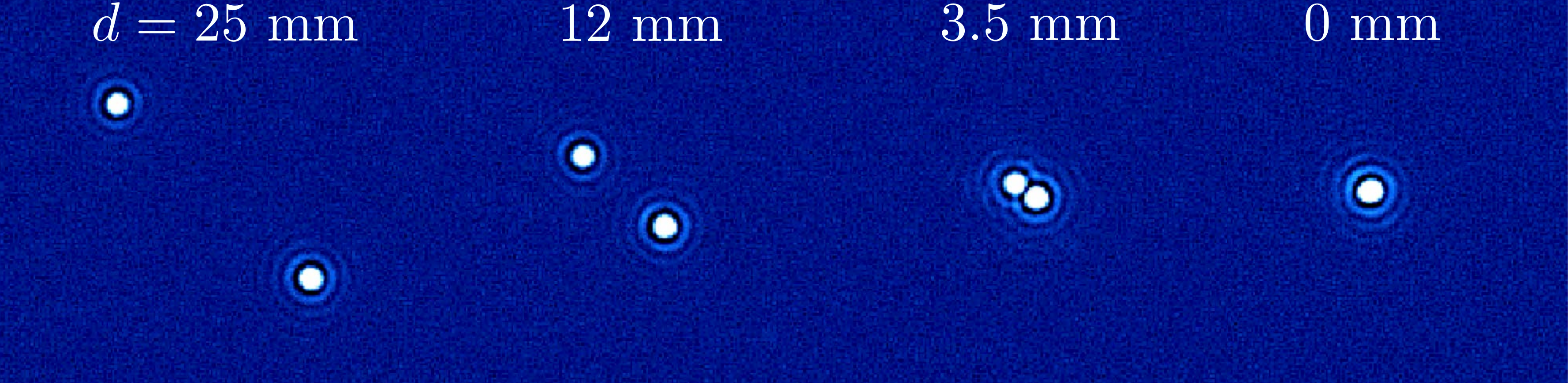}
		\caption{Lab images of the separation of the left- and right-circularly polarized beams when two polarization gratings are separated by a distance of 25, 12, 3.5 and 0 mm. 
		}
	     	\label{fig:sep_PSF}
\end{figure*}
\begin{figure*}[t]
	\centering
	\includegraphics[width= \textwidth]{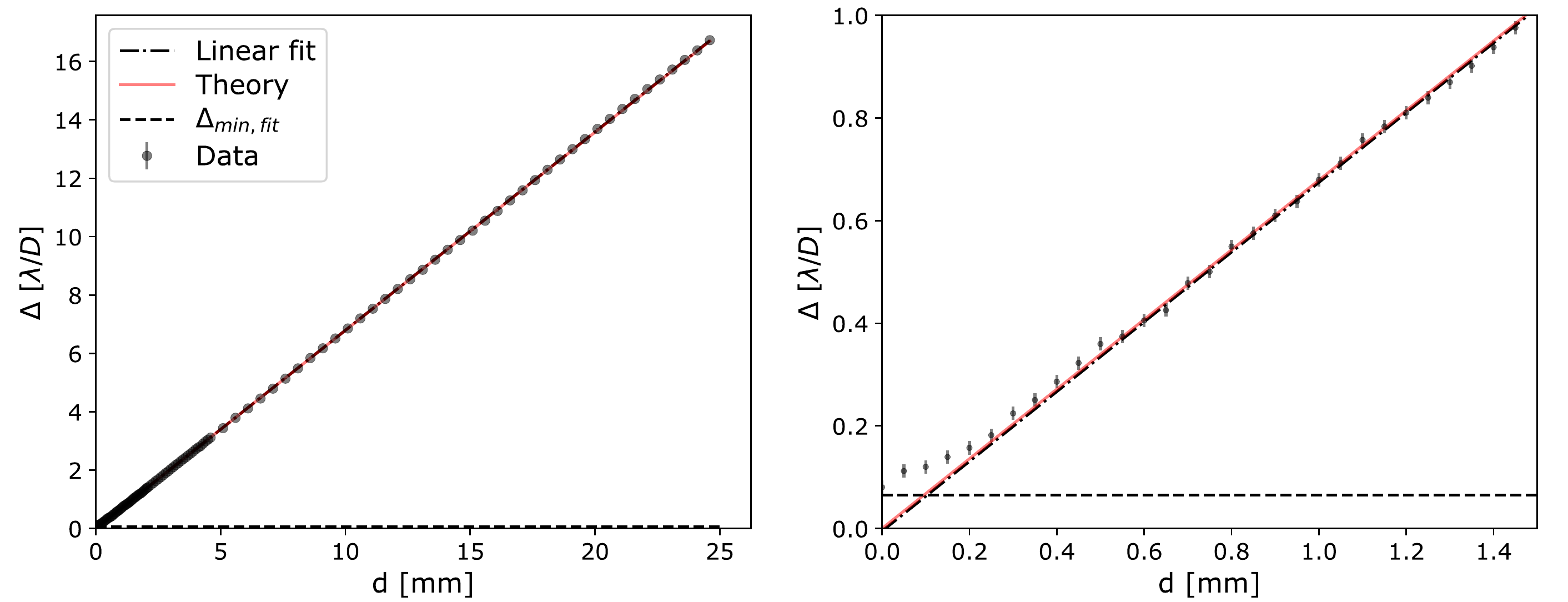}
		\caption{\textcolor{black}{\textbf{Left:} PSF separation measured as function of grating separation, with $1\sigma$ error bars. \textbf{Right:} The measured separations at the closest distances between the gratings. The data and the linear fit to the data follow the theoretical to a very high degree. The dashed line shows the minimal separation the two-PSF model can distinguish when fitted to a single PSF. }}
	     	\label{fig:sep_dis}
\end{figure*}

The distance between the gratings is controlled with a translation stage. 
We start with a separation of zero mm, with the two gratings gently pressed together. 
We take 10 images that are background-subtracted and median-combined. 
The sampling between 0 mm and 25 mm separation changes and has three different step sizes.
Between 0 and 2 mm, 2-5 mm, 5-25 mm, the step sizes are 50 micron, 100 micron, and 500 micron respectively. 
The resulting images are shown in Fig.~\ref{fig:sep_PSF} for separations of 25, 12, 3.5 and 0 mm. 
When pressed together, the two PSFs overlap to a very high degree. 
To quantify the PSF separations $\Delta$ as a function of plate separation, we model both PSFs using a non-aberrated Airy pattern and fit these to the data simultaneously. 
We plot the measured $\Delta$ as function of separation in $\lambda/D$ in Fig.~\ref{fig:sep_dis}. 
The 1$\sigma$ error bars of the data are calculated from the jitter of the center of the PSFs. 
Using the plate scale, we convert the calculated separation from Eq.~\ref{Eq:sep_v_sep} to $\lambda/D$.

The calculated and measured separation are in good agreement with each other beyond 0.5 mm separation. However, there are hints of a modulation on the order of $0.5 \lambda/D$, which might be caused by interference between the two PSFs.
To confirm the slope, we fit a line through the data and recover the slope of $0.6795 \pm 0.0003$ $(\lambda/D)/$mm ($1\sigma$) and an intercept of $-0.005 \pm 0.003$ $\lambda/D$. 
The fitted slope is not within $1\sigma$ of the theoretical slope of 0.6788 $\lambda/D/$mm. 
One explanation for the inconsistency between the data and the theory could be that the plates were not perfectly parallel, or that the grating constants of the two PGs are not exactly the same.
When the gratings are closer, the determined PSF separation $\Delta$ asymptotically reaches $0.1 \lambda/D$. 
This is most likely an artifact of the model fitting that cannot distinguish the two separate PSFs anymore for small separations.
Furthermore, the model fitting is affected by  interference between the two PSFs.
We test the model fitting sensitivity by fitting the two PSFs to a single PSF of one circular polarization state for the 30 largest separations. 
The separations are larger than 0 and the median value, $0.065 \lambda/D$, is shown in Fig.~\ref{fig:sep_dis}  as a dashed line. 
It is therefore unclear if the the theoretical separation of exactly zero is actually achieved. 

\section{Experimental results for a double-grating Vector Vortex Coronagraph}\label{Sec:DG_VVC}
\begin{figure*}[t]
	\centering
	\includegraphics[width= \textwidth]{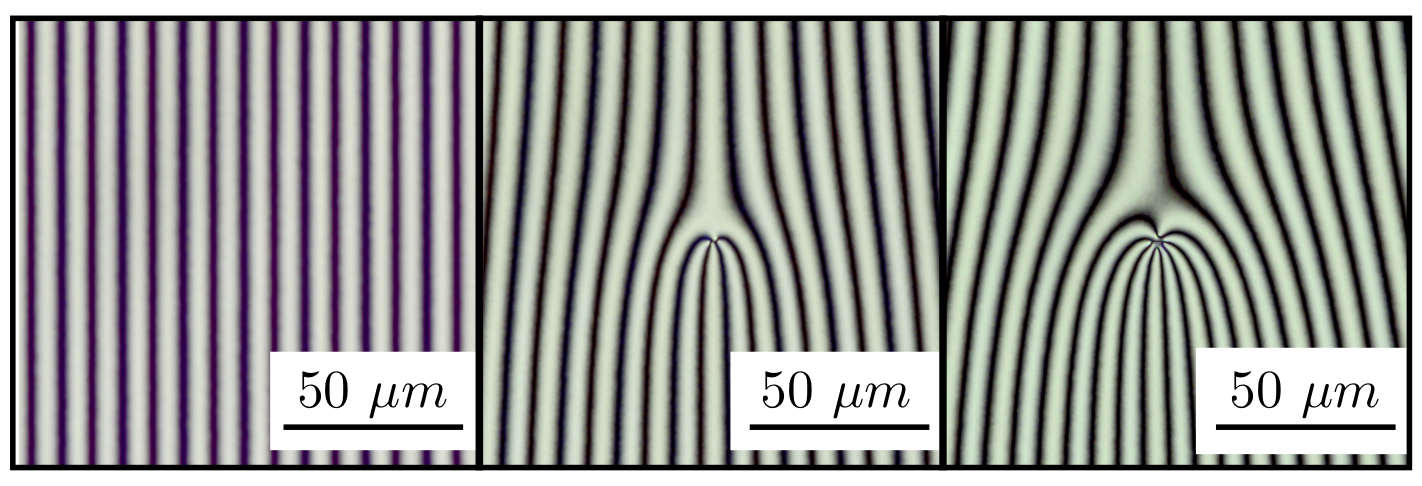}
		\caption{Microscopic images through crossed polarizers of the polarization grating and VVC + grating pattern charge 2 and 4.
		}
	     	\label{fig:vvcdemo}
\end{figure*}

We have shown that a combination of two polarization gratings can be used to reduce on-axis polarization leakage.
In this section we show that the double-grating concept can be used to reduce the impact of polarization leakage on a Vector Vortex Coronagraph (VVC).
The first polarization grating used in previous experiments is on the same substrate as a charge-2 and charge-4 vortex pattern with an added grating, i.e.~a forked grating.
\begin{figure*}[t]
	\centering
	\includegraphics[width= \textwidth]{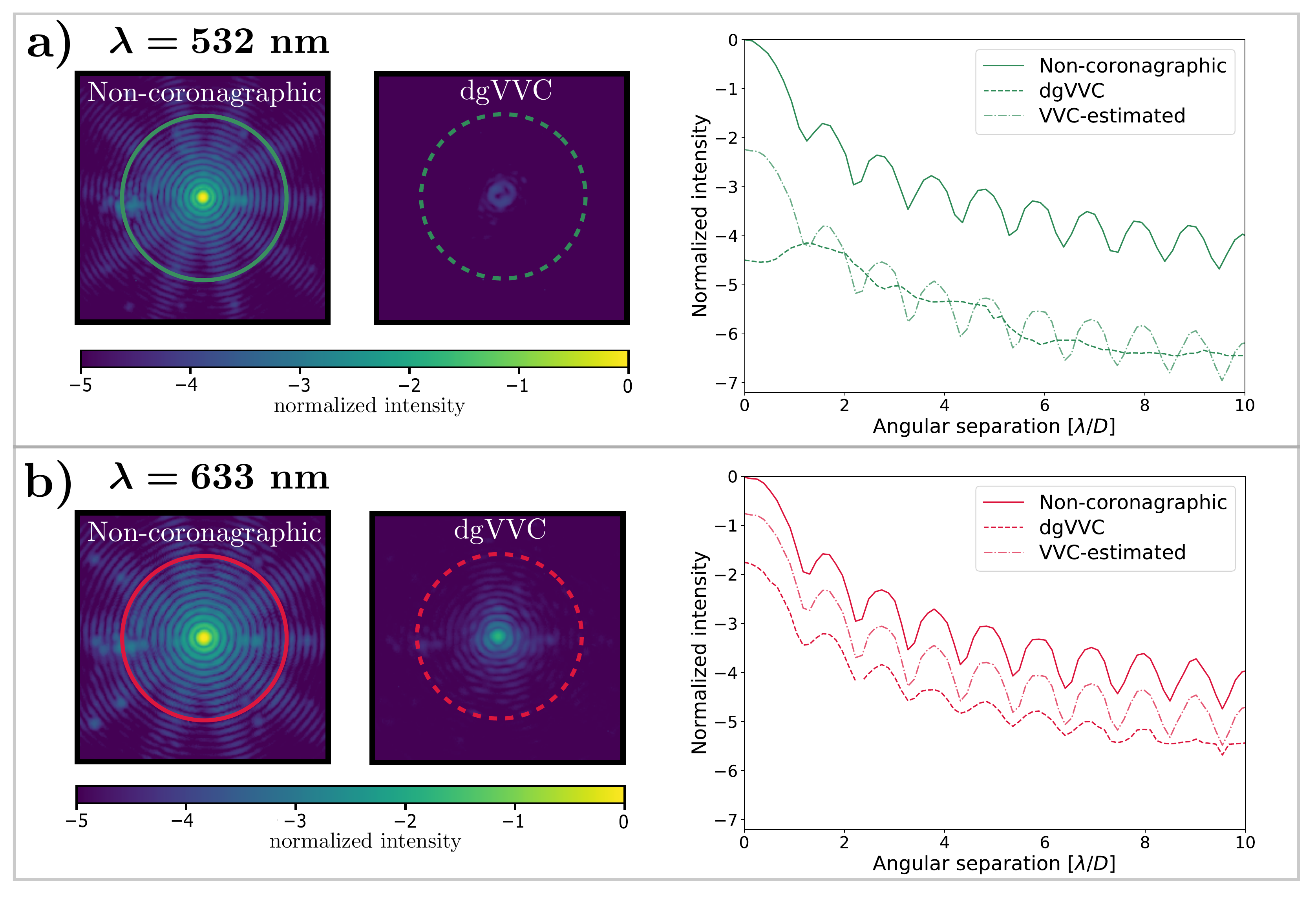}
		\caption{Measured coronagraphic and non-coronagraphic point-spread functions for the charge 4 double-grating VVC (left) and their azimuthally-averaged normalized intensities (right). We added an estimated leakage for a standard VVC (no gratings) as a reference. This VVC leakage is the non-coronagraphic PSF scaled to the level of the leakage of PG1, measured in Sect. \ref{sec:dgleakage}. 
		}
	     	\label{fig:vvccontrast}
\end{figure*}
Therefore, these coronagraphs have the same spectral retardance profile as the first grating in the experiments in Fig.~\ref{fig:DGleakagewl}. 
We image the two vortices and the second grating under a microscope between crossed polarizers.
The results are shown in Fig.~\ref{fig:vvcdemo}. 
Both the charge-2 and charge-4 vortex have a small central defect of $<1$ $\mu$m and 2 $\mu$m, respectively, and no other defects are observed. 
We conclude that the produced VVCs are of high quality. \\
We test the performance of the double-grating VVC charge 4 in the setup using configuration b shown in Fig.~\ref{fig:sep_setup}, with a Lyot stop at P2. 
From the leakage intensity measurements it is expected that the polarization leakage for 532 nm at the $10^{-5}$ level will not limit the coronagraphic performance, given the limited optical performance of our setup.
We measure the on-axis suppression, i.e.~the coronagraphic PSF, and off-axis point spread function with a grating separation of zero. 
The off-axis PSF is shifted by $>10 \lambda/D$, so it is a good approximation for a non-coronagraphic PSF. 
We conduct these on-axis and off-axis measurements at 532 nm and 633 nm. 
Note that this is a passive setup, there is no active element implemented that measures and corrects the wavefront and the performance is limited by aberrations in the system.\\
The results are shown in Fig.~\ref{fig:vvccontrast}. 
For 532 nm, we reach a contrast of $10^{-5}$ at $2.5\lambda/D$, which is in good agreement with other VVC experiments with a passive setup using polarization filtering \citep{mawet2009vector,delacroix2013laboratory}.
To compare this result with a standard VVC, we use the measured leakage of PG1 of $6\times10^{-3}$ at 532 nm, which has been manufactured on the same substrate as the coronagraphic polarization grating. 
From this comparison it seems that the gain of using a double-grating is minimal.
However, the contrast is no longer limited by leakage, which can not be corrected with adaptive optics, so adding active wavefront control would improve the contrast.
Furthermore, there is a factor 2 gain in throughput as we do not filter one of the two circular polarization states before or after the VVC mask.
Lastly, the leakage term that is used for comparison is a factor 2-4 lower than reported broadband liquid-crystal recipes \citep{Otten2017,Doelman2017}.\\
At 633 nm, the dgVVC coronagraphic PSF is dominated by the leakage term at the predicted $10^{-2}$ level. 
However, the this is an order of magnitude lower than the estimated leakage of the VVC. 
This clearly demonstrates that the double-grating element suppresses leakage.
In addition, it demonstrates that the gain of using a double-grating is only as good as the leakage of the individual liquid-crystal recipes. 
In this case, the single layer recipe provides only good leakage suppression around 532 nm. \\
\begin{figure*}[t]
	\centering
	\includegraphics[width= \textwidth]{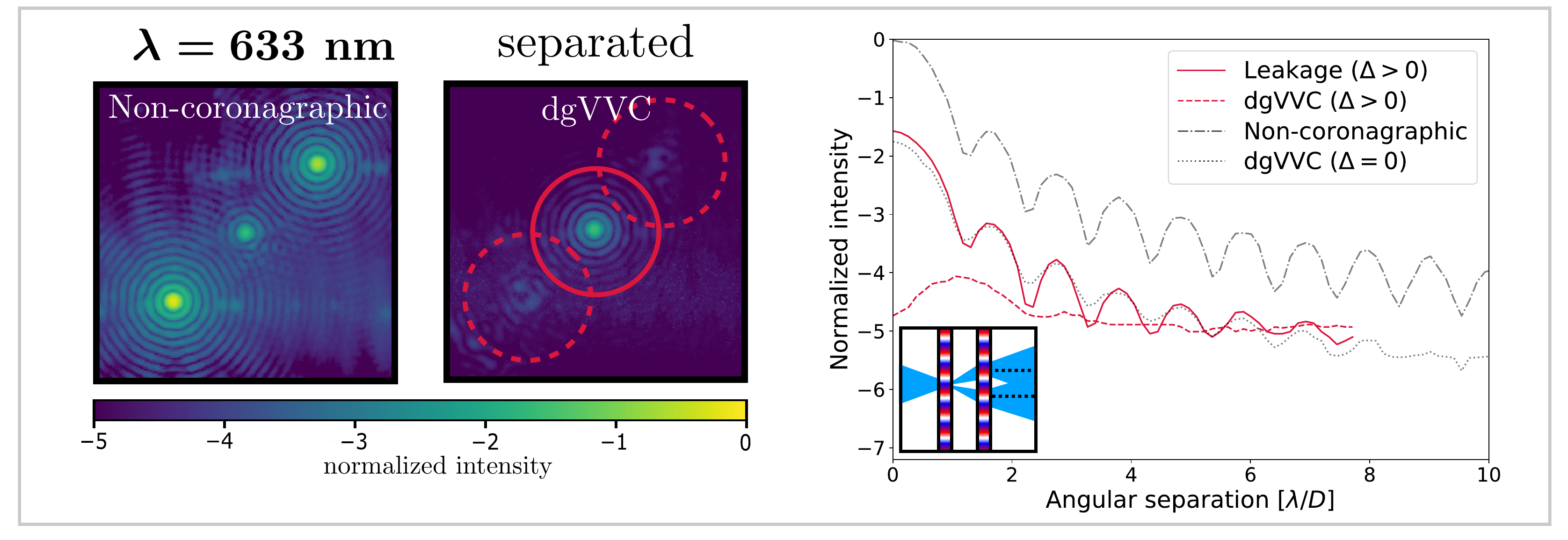}
		\caption{Measured coronagraphic and non-coronagraphic point-spread functions for the charge 4 double-grating VVC  (left), with the substrates separated by 20 mm, and the azimuthally-averaged normalized intensities (right). In gray are the non-separated measurements, as shown before in the previous figure. The non-zero distance between the substrates separates the coronagraphic and leakage PSFs on the detector. The two off-axis (coronagraphic) dgVVC radial profiles are averaged. 
		}     	\label{fig:vvcsplitcontrast}
\end{figure*}
To demonstrate that improving the liquid-crystal recipe would improve the contrast, we separate the gratings by a distance of 20 mm. 
Equivalent to Sect. \ref{sec:dgleakage}, this separates the leakage PSF from the coronagraphic PSFs. 
The resulting PSF and azimuthally averaged intensities are shown in Fig. \ref{fig:vvcsplitcontrast}.
For comparison, we added the results from in Fig. \ref{fig:vvccontrast}b) in gray.
The comparison shows that the measured on-axis coronagraphic PSF in the non-separated case, dgVVC $\Delta=0$ has a similar intensity as the leakage term for $\Delta>0$.
The difference \textcolor{black}{could} be explained by the interference between the vortex PSF and the leakage PSF.
By looking at the separated coronagraphic PSFs in \textcolor{black}{Fig.~\ref{fig:vvcsplitcontrast}}, we find that the performance of the coronagraph at 633 nm without leakage is comparable to the suppression at 532 nm. 
This shows that currently the contrast of the dgVVC at 633 nm is limited by the leakage, which is already an order of magnitude lower than for a single-element. 
We summarize that a further improved leakage suppression at 633nm would enhance coronagraphic performance.
A double-grating vector-vortex with two multi-layered liquid-crystal recipes can achieve good performance for a larger bandwidth. 
Assuming the $1-2.5\mu m$ 3TR recipe shown in Fig.\ref{fig:DGleakagebb} (green) a double-grating VVC \textcolor{black}{could} suppress this leakage to the measured level of $10^{-5}$ for $\sim$90$\%$ bandwidth. 
In conclusion, these lab results demonstrate that the double-grating concept works for the VVC and may be key to manufacturing a high-performance broadband VVC mask. 

\section{The double-grating vector-Apodizing Phase Plate for the Large Binocular Telescope.}\label{Sec:DG_LBT}

\begin{figure}[t]
    \centering
    \includegraphics[width = \linewidth]{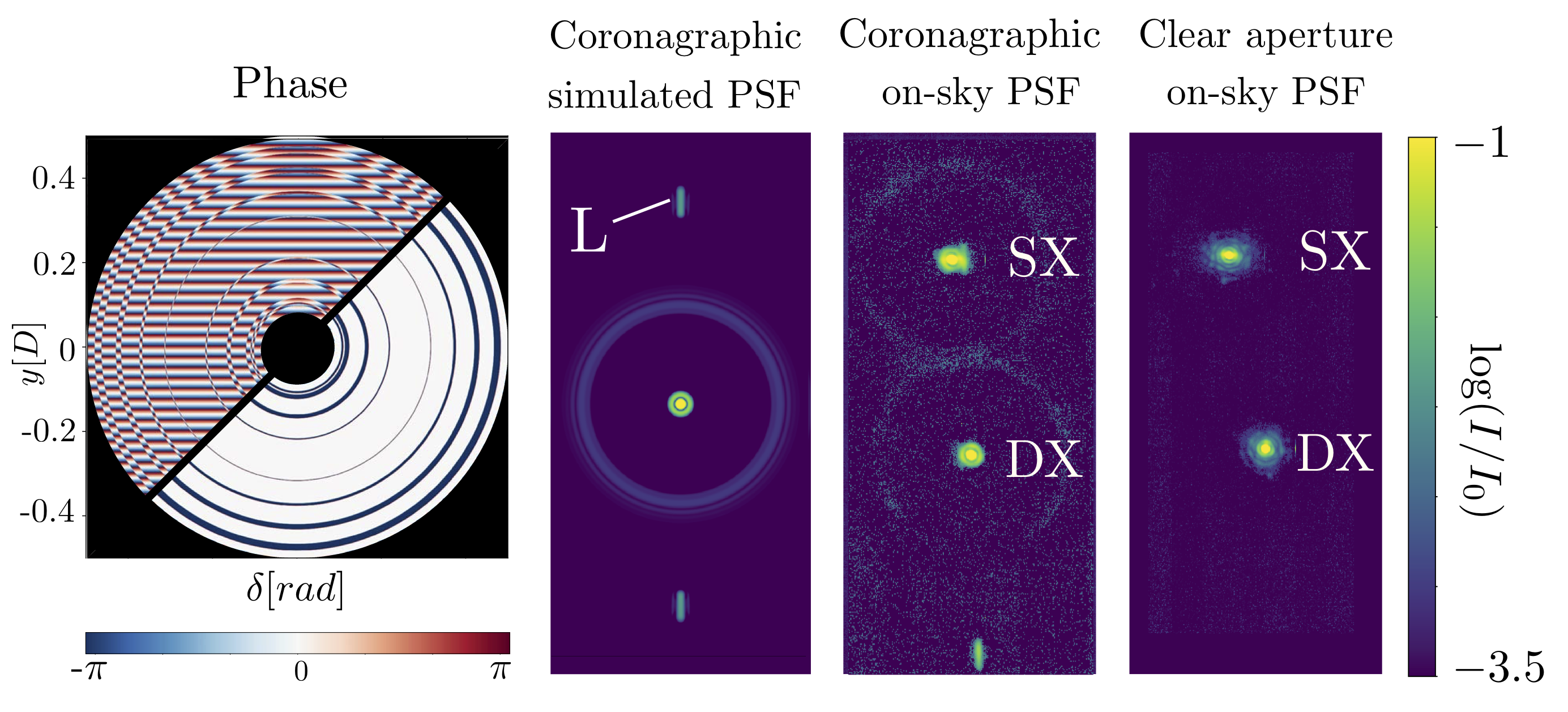}
    \caption{\textbf{Left:} vAPP phase pattern with and without grating. \textbf{Center:} Simulated broadband PSF of the vAPP, showing the diffracted leakage (L). \textbf{Right:} First light of the double-grating vAPP at LBT, showing the PSFs of both telescope apertures, SX and DX in L-band. \textcolor{black}{Note that the PSFs can be positioned such that the leakage terms are not imaged in the dark zone of the other PSF.}}
    \label{fig:PhaseDesign}
\end{figure}

In this section, we present the design and the first on-sky images of the double-grating vector-Apodizing Phase Plate coronagraph for the L/M-band (3-5 $\mu$m) InfraRed Camera (LMIRcam) \citep{skrutskie2010large} at the Large Binocular Telescope (LBT).
The design of the phase pattern of the double-grating vAPP for the LBT aims to minimize the stellar diffraction halo by generating an annular dark zone that extends from close to the core of the stellar PSF, to beyond the control radius of the deformable secondary mirror \textcolor{black}{($\sim13 \lambda/D$)}. 
This double-grating vAPP is compatible with the ALES integral-field spectrograph \citep{skemer2015first,skemer2018ales}, as its single combined PSF well matches its small field-of-view, and it offers stable coronagraphic performance over the entire wavelength range, even extending down to K-band at 2 $\mu$m. \textcolor{black}{This is enabled by the liquid-crystal recipe described in \citep{Doelman2017}}.
The design of the vAPP phase pattern is computed using the global optimizer described in Por (2017) \citep{por2017optimal} and consists of annular rings with $0$ or $\pi$ phase with an added grating with 40 cycles over the pupil. 
Both the phase pattern with and without grating are shown in Fig.~\ref{fig:PhaseDesign}. 
This design has an annular dark zone from 2.7 to 15 $\lambda/D$ with a raw contrast of $10^{-4}$ close to the inner working angle and $10^{-5}$ further out.
The inherent Strehl ratio for the coronagraphic PSF core is 46\%.

\begin{table}[]
    \caption{Fitted intensity and leakage values for L-band for both vAPPs and a clear aperture. \textcolor{black}{The intensity and leakage are fitted using least-squares fitting with unaberrated broadband PSF models to the clear aperture data and vAPP PSF models to the vAPP data separately.}}
    \centering
\begin{tabular}{c c c c c c c}
   \hline
   & Clear & Clear & dgvAPP  & dgvAPP  \\
   Telescope aperture & DX & SX & DX & SX \\
   \hline
  $I_{Norm}$ & 0.81 & 1.0 &  0.44 & 0.45 \\
  Diffracted leakage & - & -& 0.04 & 0.01\\    
  Estimated zero-order leakage & - & -& $1.6 \times 10^{-3}$ & $1 \times 10^{-4} $\\
  \hline
\end{tabular}

  \label{tab:LBT_fit_param}
\end{table}
The presented design for the double-grating vAPP has been installed inside LMIRCam on the Large Binocular Telescope early September 2018. 
The first on-sky results have been obtained on \textcolor{black}{December 24th 2018, with the aim of demonstrating the double-grating concept on-sky. 
The star HIP 75097 was observed in L-band (std-L, 3.41 $\mu m$ - 3.99$\mu m$) with and without the double-grating vAPP coronagraph with 1.3 arcseconds seeing at an airmass of 1.36. The total integration time of with the vAPP is 5.5 seconds and the total integration time with a clear aperture is 25 seconds.}
The reduced image, containing PSFs for both telescope apertures are shown in Fig.~\ref{fig:PhaseDesign}. 
For comparison, we simulate the PSF of a single aperture over $12\%$ bandwidth using HCIPy \citep{por2018hcipy}. 
The simulations visually match the on-sky data apart from the AO residuals (mostly wind-driven halo) and low-order aberrations.
Because only one PSF (including diffracted leakage term PSFs) is visible for each telescope aperture, it can be concluded that the PSFs of both polarization states overlap and the double-grating principle works, and the manufacturing and installation was successful. \\
We investigate the on-sky transmission and leakage from the L-band data by fitting PSF models to both the coronagraphic and non-coronagraphic data using least-squares fitting. 
The bandwidth of the models is 3.55 $\mu$m to 3.95 $\mu$m, where the throughput is modelled as a tophat profile. 
The fit parameters are the position on the detector, the plate scale, the PSF intensity and the residual background. 
For the vAPPs we also fit phase pattern rotation, leakage and the circular polarization fraction (derived from the flux ratio of the leakage PSFs) for each telescope aperture seperately. 
The fitted intensities and leakage values can be found in Table \ref{tab:LBT_fit_param}. 
Here $I_{norm}$ is the transmission normalized on one of the two PSFs ("SX") for a clear aperture, the diffracted leakage is the intensity of the leakage terms obtained from the off-axis PSFs, and the on-axis leakage is estimated from these off-axis leakage terms. 
Note that the inherent Strehl is not included in the transmission. 
The transmission below 50\% can be explained by a strong absorption feature of the glue (Norland NOA 61) and liquid crystals between 3.55 $\mu m$ to 3.7 $\mu$m.
Furthermore, both apertures (``SX'' and ``DX'') do not have the same transmission, where DX has a transmission that is 20\% lower than the SX aperture and the PSF is severely aberrated. 
For the vAPP no such effect is seen, where $I_{norm}$ is the same for SX and DX.
This could be explained with a slight pupil misalignment for the clear aperture.
Because the vAPP apertures are undersized, the misalignment has a smaller impact on the vAPP.

 \begin{figure*}[t]
	\centering
	\includegraphics[width=\textwidth]{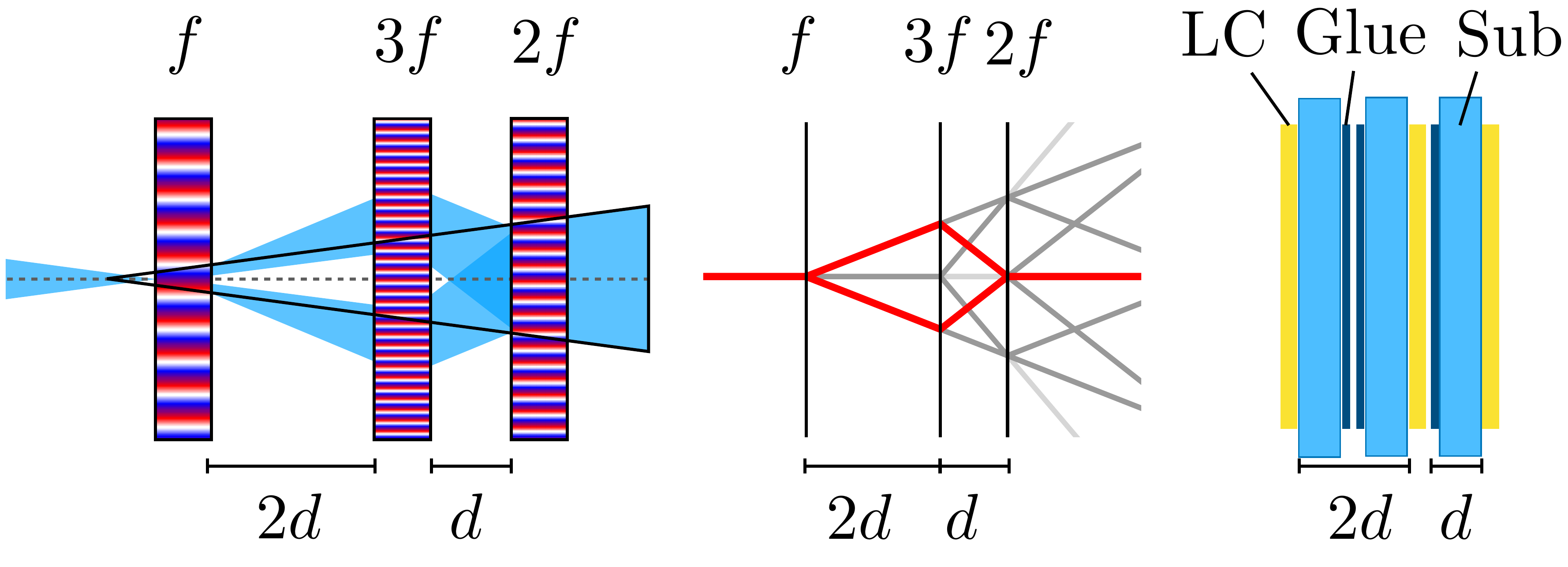}
		\caption{\textbf{Left:} \textcolor{black}{Schematic of the triple grating concept. By combining three polarization gratings with different grating frequencies, f, in a 1:3:2 ratio, it is possible to overlap the beams with opposite circular polarization.}  This additional path length results in a shifted virtual focus. \textbf{Center:} Schematic of the paths of all diffracted terms for the triple grating. The gray scale indicates the intensity of the brightest leakage term. The red lines indicate the path of the beams that are diffracted by all gratings. \textbf{Right:} Implementation of the triple-grating concept with single thickness substrates.}
	     	\label{fig:tgschematic}
\end{figure*}

\section{Multi-grating coronagraphs and system-level perspectives}\label{sec:systemimplications}

\textcolor{black}{We can extend the multi-grating principle from the double-grating VVC in Sect.~\ref{Sec:DG_VVC} to a triple-grating or generally $N$-grating implementation.
The additional layer(s) of leakage-term diffraction enable better intrinsic contrast performance and/or achieving that contrast over larger spectral bandwidths.
Moreover, these techniques can avoid PSF splitting \citep{Komanduri2011}.
First we will show the triple-grating concept and expand this to $N$ gratings.}

\subsection{Multiple grating combinations}\label{Sec:multi-grating}
The schematic of the triple grating element is shown in Fig.~\ref{fig:tgschematic}.
The triple-grating element consists of three gratings with different periods in a 1:3:2 ratio.
The primary and the conjugate beams of the first grating are diffracted in the opposite direction by the second grating and combined by the third grating. 
Only one leakage wave and the primary and the conjugate beams can end up on-axis, i.e.~the triple leakage and the triply diffracted beams. 
No other combination of the grating phases adds up to zero. 
To recombine both circular polarization states at the optical axis, it is required that the distance ratio between the gratings is exactly 1:2. 
This can be achieved with high accuracy if the liquid-crystal films and their substrates are positioned as shown in the right in Fig.~\ref{fig:tgschematic}. Any deviation from the 1:2 ratio will separate the PSFs according to Eq. \ref{Eq:sep_v_sep}.  
Note that for the pure grating patterns, only their orientation is relevant in the mutual alignment.
With an average polarization leakage of $1\%$ for a single grating, the combined leakage would be $10^{-6}$ at the center, and $10^{-8}$ at the first Airy ring.. 
Such a triple-grating (or any other $N$-grating) configuration in the focal plane does lead to a minor shift in the (virtual) focus.\\
For multi-element combinations, i.e. $N>2$, there are two constraints. 
First, the total sum of the grating frequencies should add up to zero, taking into account the switch of the sign of the polarization state with each element, i.e. $\sum_i^N (-1)^i f_i = 0$. 
Only then the beam that is diffracted by all elements is parallel to the optical axis.  
Second, all the grating frequencies should be unique. 
This ensures that no other combination of frequencies is zero. 
Note that these frequencies can be negative, such that the grating direction is inverted. 
These constraints produce solutions that are scale invariant, i.e.~they only constrain the ratios of the grating frequencies and distances. 
Hence, we are free to normalize the grating frequencies on the minimum frequency.
There are multiple solutions for a single $N$-grating combination for $N>3$. 
To ensure manufacturability the solutions should ensure the lowest grating frequency.

\subsection{Applications of multiple-grating focal-plane coronagraphs}\label{Sec:multi-grating-applications}
\textcolor{black}{
For space applications, the double-grating and the triple-grating VVC implementations offer exciting opportunities to deliver extremely high intrinsic contrasts over large spectral bandwidths.
We already showed that in simulation a multi-layered double-grating elements is able to suppress leakage to $10^{-10}$ for bandwidths up to $20\%$, see Fig.~\ref{fig:DGleakagebb}(d).
We expand these simulations to a triple-grating element, optimizing the liquid-crystal recipes for three PGs simultaneously. 
These simulations show that a triple-grating VVC implementation has the potential to reach $10^{-10}$ leakage suppression on the first Airy ring.
The optimized triple-grating element suppresses the leakage-term by eight orders of magnitude over the entire visible wavelength range, without any external polarization filtering, see Fig.~\ref{fig:tgleakage}.
\begin{figure}[t]
    \centering
    \includegraphics[width = \linewidth]{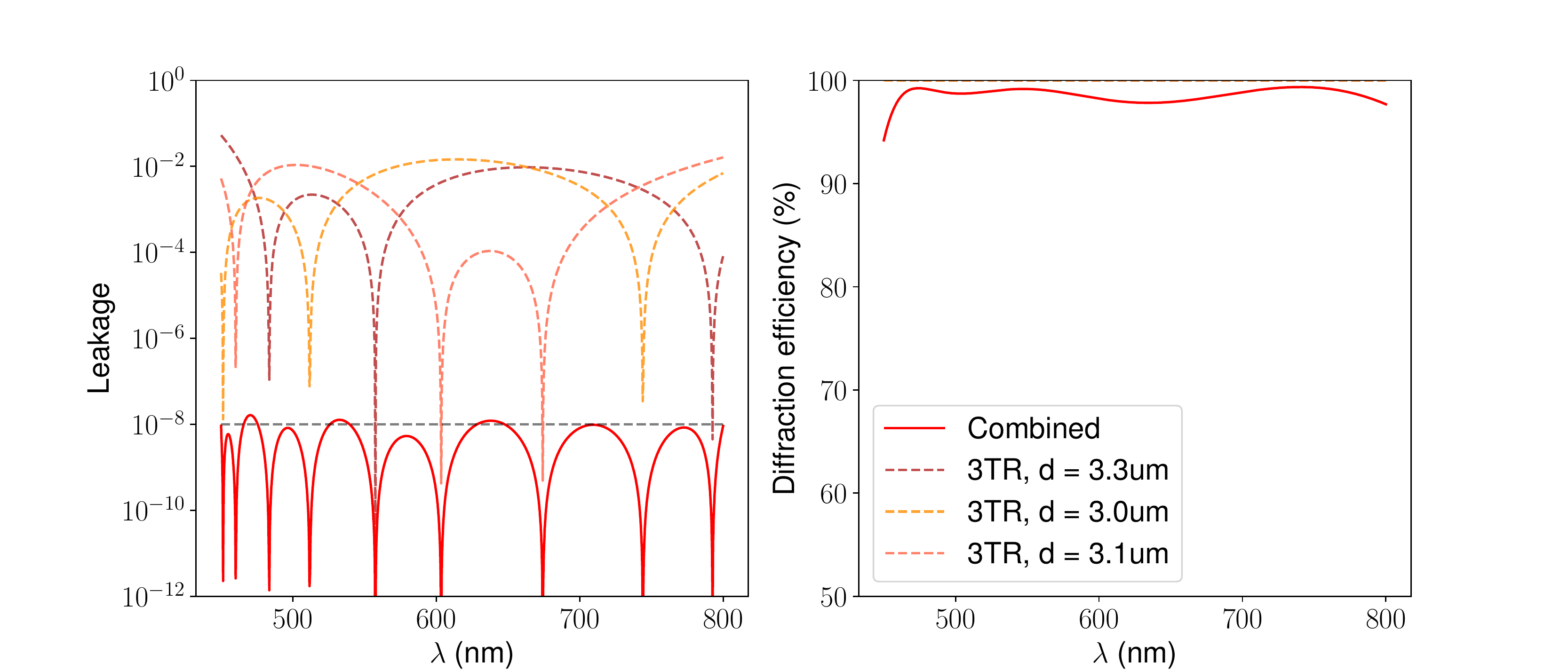}
    \caption{\textbf{Left:} Simulated leakage as function of wavelength for a triple-grating VVC in the visible wavelength range (450-800 nm). Each element is a 3TR with a slightly different recipe. \textbf{Right:} Simulated diffraction efficiencies for the same triple-grating VVC. The average diffraction efficiency is 98\%.}
    \label{fig:tgleakage}
\end{figure}
Moreover, a tgVVC has in theory no split-up of the off-axis PSF, as a dgVVC may have to a small extent.
In future work, we will perform detailed simulations of the performance of a tgVVC6 (including full internal ray-tracing), and validate these simulations with a prototype element at a high-contrast test-bed.}\\
\textcolor{black}{In future space missions dedicated to high-contrast imaging, like the HabEx concept \citep{HabEx2019} that features a 4-m off-axis telescope that is ideal for a charge-6 Vortex coronagraph, a tgVVC can have a major impact on the system-level design and trade-offs:\\
A single tgVVC has the potential to suppress the leakage to $10^{-10}$ on the first Airy ring in any 20\% band over the entire visible range, meaning that a dichroic split-up in e.g.~a blue arm and a red arm would be no longer necessary.
Two identical arms can then implement filter wheels that contain all relevant filters, to still perform simultaneous observations in complementary spectral bands, or in the same spectral band with complementary dark-hole shapes.
This architecture provides full redundancy in case of a single-point failure in either arm.
Moreover, it can be considered to perform spectroscopic observations over a larger spectral band than 20\%, and rely on a single-mode fiber feed and spectral differential techniques to compensate for the intrinsically worse contrast performance of the broadband system \citep{PorHaffertSCAR,Coker2019}.}\\
In theory, it is not required to perform any polarization filtering when using the tgVVC6, and all light could go through. 
The charge-6 VVC has been selected for its robustness to low-order aberrations like tip/tilt and astigmatism, and particularly their polarization-dependent versions that invariably emerge for converging beams on fold mirrors and telescopic systems, respectively \citep{Davis2018}.
In practice, however, it may still be necessary to install polarization optics.
This would then mostly serve to mitigate the effects of polarization aberrations \citep{Breckinridge2015} and facilitate polarization-dependent dark-hole control.
Filtering linear polarization can then be combined with converting the linear polarization state to circular polarization, ensuring that only of the of two vector vortex twist modes, as induced by the geometric phase, is controlled by the AO system.
Note that no further polarization filtering optics are then required after the tgVVC6 mask, which alleviates the overall instrument complexity.
For now, we assume that a single layer of linear polarization filtering plus conversion to circular polarization is part of a realistic systems implementation.\\
We can combine the aforementioned strategies.
With a two-arm tgVVC6 implementation, linear polarization filtering can be implemented with a polarizing beam-splitter that feeds the two identical arms, instead of a dichroic splitter \citep{HabEx2019}.
This configuration now allows the implementation of dual-beam polarimetry with identical spectral bands and dark holes in both arms.
This system can be upgraded to a complete linear beam-exchange polarimeter by including a rotating half-wave plate in front of the polarizing beam-splitter, polarization aberrations permitting \citep{SnikKellerreview,Snik2014a}.

\section{Conclusions}

We conclude that multi-grating geometric phase hologram elements, consisting of two or more polarization gratings, can be used to generically suppress polarization leakage of diffractive phase plate coronagraphs by orders of magnitudes compared to a single element.
We demonstrate that by adding a coronagraphic phase pattern to the polarization grating of the first element, it is possible to make low-leakage geometric phase coronagraphs, i.e.~a Vector Vortex Coronagraph and a vector-Apodizing Phase Plate coronagraph. 
We show in the lab that the double-grating Vector Vortex Coronagraph has much reduced on-axis leakage and has similar performance as other VVC coronagraphs in literature. 
We have demonstrated the double-grating concept on-sky with a newly installed double-grating vector-apodizing phase plate (vAPP) coronagraph, operating in LMIRCam at the Large Binocular Telescope.
For ground-based telescopes, double-grating coronagraphs like the double-grating VVC, preferably of charge 2 (dgVVC2), and the dgvAPP360 are ideally suited to work in tandem with integral-field spectrometers, as they offer a single PSF that easily fits within the field-of-view, and, moreover, delivers stable, high-contrast performance over spectral bandwidths as large as a full octave.
We aim to use our double-grating vAPP over the full bandwidth of the ALES integral field spectrograph in LMIRCam to detect and characterize exoplanets in K, L \& M band. 
Furthermore, we will install a 1-2.5 $\mu$m double-grating VVC in the SCExAO instrument at the Subaru telescope to feed the CHARIS instrument integral field spectrograph. 
Finally, we will further develop and analyze double-grating and triple-grating VVCs that are optimized for $\sim$10$^{-10}$ raw contrast of future space instrumentation, and demonstrate contrast performance over large spectral bandwidths.

\section*{Funding}
ERC Starting Grant 678194 (FALCONER);

\section*{Acknowledgments}
Part of this work was carried out at the Jet Propulsion Laboratory, California Institute of Technology, under contract with the National Aeronautics and Space Administration (NASA).
We thank our all our collaborators from the LBT for helping us acquire the first on-sky results with the double-grating vAPP, especially Phil Hinz, Steve Ertel, Jordan Stone, and Eckhart Arthur Spalding.
This research made use of HCIPy \citep{por2018hcipy}, an open-source object-oriented framework written in Python for performing end-to-end simulations of high-contrast imaging instruments.
We thank Christoph Keller for the many
fruitful discussions, which helped improve the results presented in this work.
We thank Matthew Kenworthy for doubting that this principle would ever work, but being a good sport when indeed it did.

\bibliography{Double_grating}
\bibliographystyle{AAS.bst}

\end{document}